\documentclass[aps,nofootinbib]{revtex4}
\input epsf
\newcommand{\sfig}[2]{\centerline{ \epsfxsize = #2 \epsfbox{#1} }}

\begin{document}

\newcommand{\Mpc}{\mbox{Mpc}}

\newcommand{\ncl}[0]{\bar{n}_{cl}}
\newcommand{\nn}{\nonumber}
\newcommand{\avgtheta}{\langle \theta|m \rangle}
\newcommand{\avgthetap}{\langle \theta|m' \rangle}
\newcommand{\avgN}{\langle N|m \rangle}
\newcommand{\NNm}{\langle N(N-1)|m \rangle}
\newcommand{\NNNm}{\langle N(N-1)(N-2)|m \rangle}

\newcommand{\eval}[0]{\mbox{\Large $\vert$\normalsize}}

\newcommand{\na}{n_{\vec a}}
\newcommand{\nap}{n_{\vec a'}}
\newcommand{\psia}{\psi_{\vec a}}
\newcommand{\psiap}{\psi_{\vec a'}}
\newcommand{\ba}{b_{\vec a}}
\newcommand{\bap}{b_{\vec a'}}
\newcommand{\Na}{N_{\vec a}}
\newcommand{\Nap}{N_{\vec a'}}
\newcommand{\ra}{r_{\vec a}}
\newcommand{\rap}{r_{\vec a'}}
\newcommand{\fa}{f_{\vec a}}
\newcommand{\fap}{f_{\vec a'}}
\newcommand{\C}{C_{\vec a,\vec a'}}
\newcommand{\nua}{\nu_{\vec a}}
\newcommand{\nuap}{\nu_{\vec a'}}
\newcommand{\da}{\delta_{\vec a,\vec a'}}
\newcommand{\xia}{\xi_{\vec a,\vec a'}}

\newcommand{\avgpsia}{\langle \psia|m \rangle}
\newcommand{\avgpsiap}{\langle \psiap|m' \rangle}

\newcommand{\tna}{\na^t}
\newcommand{\tnap}{\nap^t}
\newcommand{\tba}{\ba^t}
\newcommand{\tbap}{\bap^t}
\newcommand{\tca}{c_{\vec a}^t}
\newcommand{\tcap}{c_{\vec a'}^t}
\newcommand{\tC}{\tilde \C}
\newcommand{\txia}{\xia^t}

\newcommand{\dom}[1]{$\Delta \Omega_m = {#1}$}
\newcommand{\dtilt}[1]{$\Delta n = {#1}$}


\title{Halo Model Analysis of Cluster Statistics}
\author{Eduardo Rozo$^{1,2}$, Scott Dodelson$^{3,4}$, Joshua A.
Frieman$^{2,3,4}$}

\affiliation{$^1$Dept. of Physics, The University of Chicago, Chicago, IL 60637}
\affiliation{$^2$Center for Cosmological Physics, Chicago, IL 60637}
\affiliation{$^3$NASA/Fermilab Astrophysics Center Fermi National Accelerator
			 Laboratory, Batavia, IL 60510}
\affiliation{$^4$Dept. of Astronomy and Astrophysics, 
		  The University of Chicago, Chicago, IL 60637}

\begin{abstract}
\vspace{0.2 in}
We use the halo model formalism to provide expressions for cluster 
abundances and bias, as well as estimates for the correlation matrix between 
these observables.  Off-diagonal elements due to scatter in the 
mass tracer scaling with mass are included, as are 
observational effects such as 
biases/scatter in the data, 
detection rates (completeness), and false detections (purity).
We apply the formalism to a hypothetical
volume limited optical survey where the cluster mass tracer is chosen to
be the number of satellite galaxies assigned to a cluster.
Such a survey
can strongly constrain $\sigma_8$ ($\Delta\sigma_8\approx 0.05$), 
the power law index $\alpha$ where 
$\langle N_{gal}|m\rangle= (m/M_1)^\alpha$
($\Delta\alpha\approx0.03$), and perhaps even the Hubble parameter
($\Delta h\approx 0.07$). We find cluster abundances and
bias are not well suited for constraining $\Omega_m$ or the amplitude $M_1$.
We also find that without bias information $\sigma_8$ and $\alpha$
are degenerate, implying constraints on the former are strongly
dependent on priors used for the latter and vice-versa.
The degeneracy stems from an intrinsic
scaling relation of the halo mass function, and hence it should be
present regardless of the mass tracer used in the survey.
\end{abstract}  

\maketitle

\section{Introduction}

Two of the simplest measures of how matter is distributed throughout the
universe are its average density, parametrized by the ratio of the
matter density to the critical density, $\Omega_m$, and its
power spectrum $P(k)$.  The amplitude of the power spectrum today
is usually characterized by $\sigma_8$,
the average rms mass fluctuation in spheres of $8h^{-1} \Mpc$.
Accurate determinations of both 
$\sigma_8$ and $\Omega_m$ are of crucial importance to cosmology
as they provide some of the simplest probes of large scale structure.

Cluster abundances are well known for their ability to constrain
both $\sigma_8$ and $\Omega_m$ (see e.g. \cite{Pierpaolietal},\cite{Viana03},
\cite{Allenetal},\cite{Ikebeetal},\cite{Pierpaoli01},\cite{VianaLiddle},
\cite{Kochanek03a},\cite{bahcall2}, \cite{Schueckeretal}
and references therein). 
This seems intuitively 
reasonable: the number of large massive objects ought to 
depend on the total mass available ($\Omega_m$) and a measure 
of how likely is it for mass to clump at cluster scales ($\sigma_8$). 
More formally, from numerical simulations 
(\cite{Jenkins},\cite{Evrard02}) 
and theoretical considerations
(\cite{PressSchechter}, \cite{ShethTormen}) we know to a reasonable accuracy
what the halo mass function looks like in various cosmologies.  Even though
mass is not directly observable, by identifying a mass tracer and its relation
to halo mass one may hope to constrain cosmology.  Some examples of mass 
tracers in clusters are X-ray temperatures and luminosities of the 
inter-cluster gas, optical luminosities, and the number of galaxies found
in the cluster.

The actual analysis of data may be quite involved.  In particular, not only
is it necessary to understand how a mass tracer scales with halo mass, 
one also needs to understand both what the uncertainties in the scaling
are and the intrinsic scatter around the mean relation between the mass
tracer and halo mass (see e.g. Pierpaoli et al. \cite{Pierpaolietal},
Viana et al. \cite{Viana03} for two recent, detailed analyses).

Here, we develop simple expressions for the number density and 
bias of clusters binned using an arbitrary mass
tracer.  While much effort has been devoted to converting observations to
the theoretically simple mass function, here we attempt to massage the theory
to fit the observations: i.e. use the halo model to make predictions for 
any experiment. We attempt to include many of the most relevant 
experimental effects, including intrinsic scatter in the mass tracer, 
scatter and/or bias
arising from experimental measurements, and imperfect detection rates and 
false detections. We feel this is important since it allows
direct comparison of theory to data: by minimizing the amount of
data manipulation, the probability of artificially biasing the data
is diminished.  

We also provide theoretical estimates of the 
correlation matrix between various bins.  Our estimates 
include Poisson noise, sample variance, 
and uncertainties due to scatter in the scaling of the mass tracer
with mass, the treatment of which we believe is new.

Following the development of our formalism, we apply it to 
a model cluster catalogue that  mimics the type of 
catalogues one can construct from large optical surveys such as 
the SDSS \cite{Yorketal} or the 2dF \cite{Collessetal}.  For concreteness,
we take the mass tracer to be the number of member galaxies in a 
cluster. This has the interesting consequence that we
can analyze both cosmological constraints and constraints on
how galaxies populate halos.

From the cosmological point of view, our
results are of importance since any new measurements of $\sigma_8$ 
may help narrow the large range of measured values for 
this quantity.  More importantly perhaps, our analysis identifies
degeneracies between cosmology and the halo occupation distribution.
Not only is this an interesting problem in itself 
(see e.g. Zheng et al. \cite{Zhengetal}, Berlind \& Weinberg 
\cite{BerlindWeinberg}), but the identification of degeneracies in our
survey suggests that, in general, there will be degeneracies between
cosmology and the mass tracer scaling relation.  Said degeneracies
may help bring into agreement seemingly conflicting results for
$\sigma_8$ obtained with different
assumptions of the mass tracer scaling relation.

From the point of view of constraining galaxy formation,
determining both cosmology and the
halo occupation distribution simultaneously is important since
it avoids possible systematic errors that may arise from the choice 
of an incorrect cosmology (again coming back to the question of 
degeneracies). 
Further, we believe
that considering cluster abundances and large scale bias has the important
advantage that neither halo profiles nor second order moments of the
halo occupation distribution appear in our formulae explicitly.
This makes our results very insensitive to said variables, thereby
providing a first stepping stone toward the full determination of the 
halo occupation distribution. In particular, this type of analysis should
complement well halo occupation constraints from galaxy clustering (see e.g.
Jing, Mo, and 
B\"orner \cite{JingMoBorner}, Scoccimarro et al. \cite{Scoccimarro}, 
Moustakas and Somerville \cite{MoustakasSomerville}, Cooray \cite{Cooray}),
as well as more sophisticated studies of halo occupancy 
such as work based on the conditional luminosity function (see 
Yang, Mo and van den Bosch \cite{Yangetal} and van den Bosch, Mo, and
Yang \cite{vandenBoschetal}).  

In section 2, we develop our formalism and find expressions for the 
number density and bias of clusters binned according to measurements
of an arbitrary mass tracer.  In section 3 we identify the various sources
of uncertainty intrinsic to observations, i.e. uncertainties that would
exist even for a perfect experiment. In section 4, we 
show how to include various experimental effects both in the predictions
for what will be observed and in the uncertainties associated with the data.
Having finished our formalism, we present in section 5 the assumptions for our
model survey and characteristics of the assumed cluster catalogue.  These
are supposed to mimic the real catalogues which one may expect to construct
with surveys such as the SDSS and 2dF.  Section 5 sets up our fiducial model
and states all assumptions used to obtain our results, which are
presented in section 6.  Section 7 addresses how the results
change if we do not have bias information and consider cluster abundances 
alone.  We present our conclusions in section 8.  Also included as 
an appendix is a more thorough discussion of  
what is usually called the cluster abundance normalization condition
($\sigma_8\Omega_m^\gamma\approx0.5$, $\gamma\approx 0.5$) than the 
one presented in the main text.

\section{Halo Model Formalism and Cluster Statistics:}

\subsection{The Halo Model Approach}

The Halo Model is a theoretical framework developed to understand clustering 
properties of different mass tracers in the universe.  The halo model does 
this by dividing the problem in two: 
first, it assumes all mass in the universe is distributed in 
units called halos. The halo model then assumes
that all properties of mass tracers within a halo (e.g. galaxies, 
X-ray temperature, etc.) are determined exclusively by the
physical properties of the parent halo (e.g. mass, angular momentum, and
so on).\footnote{In the simplest cases, halos are taken to be spherical 
distributions of dark matter, with some specified density profile, typically 
an NFW \cite{NFW} or Moore \cite{Moore} profile.  For our purposes, neither 
the shape nor the mass distribution of the halos will be important.}

Let then $\eta$ be our mass tracer, e.g. X-ray temperature/luminosity, 
optical luminosity, or number of member galaxies.
We will be interested in the clustering
properties of halos as a function of the tracer $\eta$. In particular, 
we will be interested in the density and bias of halos for
an arbitrary binning criterion $\psi(\eta)$.  For instance,
one may wish to bin clusters by specifying the maximum
and minimum values $\eta$ may take for a cluster to be included
in a specific bin.  This corresponds to a top-hat selection function
$\psi(\eta)=1$ when $\eta_{max}>\eta\geq\eta_{min}$ 
and $\psi(\eta)=0$ otherwise. Since we will be interested in medium and
large mass halos, we will be using the terms halos, groups, and 
clusters interchangeably.

\subsection{Cluster Density}

Let us write then  the cluster density in the halo model formalism.  
Let the $i^{th}$ halo be located at position $\vec x_i$ and let
$\eta_i$ be the value of the mass tracer $\eta$ for that particular halo.
Then the density of objects where $\eta$ is larger than some specified
minimum value $\eta_{min}$ is given by
\begin{equation}
n_{cl} (\vec x) = \sum_i \delta (\vec x -\vec x_i)\theta (\eta_i-\eta_{min} )
\end{equation}
where the sum is over all halos. Here, 
$\theta(x)$ is the usual step function, i.e. $\theta (x)=1$ for $x\geq 0$ 
and $\theta (x)=0$ otherwise, so that a halo contributes to 
the density if and only if $\eta \geq \eta_{min}$.
Alternatively, one may be interested in some other selection criteria,
e.g. looking at objects with $\eta_{max} > \eta \geq \eta_{min}$,
characterized by a top-hat function as discussed above.
Let $\psia (\eta)$ represent an arbitrary window function used to bin
data, where $\vec a$ contains the parameters that specify the binning.  
In this case, the above expression becomes
\begin{equation}
\tna(\vec x)= \sum_i \delta (\vec x -\vec x_i) \psia(\eta_i)
\end{equation}
where the superscript $^t$ is meant to signify that this is the true 
density of objects in our bin: i.e what we would observe with perfect 
instruments.  Systematic effects brought about by observations may lead
to an observed density $\na \neq \tna$. We fold these into our 
formalism in \S 4.

We now make use of the ergodic hypothesis, and assume that the
spatial clustering statistics of any field are identical to those obtained
upon averaging the corresponding expressions over a hypothetical ensemble
of universes.  Using $\langle \rangle$ to denote ensemble average, and
an over bar $\mbox{ } \bar{} \mbox{ }$ to denote spatial averaging,
we expect the spatially averaged cluster density to be:
\begin{equation}
\bar \tna = \langle \sum_i \delta (\vec x -\vec x_i) \psia (\eta_i)
	       \rangle
\end{equation}

In order to obtain expectation values, one needs to know then the
probability distribution for $\eta_i$.  We make the standard assumption
that the value of the observable $\eta$ for a halo of mass $m$ is a random 
variable with a probability distribution $P(\eta|m)$ depending
only on the mass $m$ of the halo.  For instance, the assumption
of hydrostatic equilibrium in clusters allows one to 
relate the observed X-ray temperature to the mass of the cluster.
Likewise, simulations seem to indicate
that the number of galaxies $N$ in a halo of mass $m$ is relatively 
insensitive to environment, so that $P(N)$ depends only on $m$
\cite{Andreas},\cite{LemsonKauffmann}.

Writing $\eta_i=\eta(m_i)$, where $\eta$ is a random variable for each 
mass value $m_i$,
we can rewrite the expression above as:
\begin{eqnarray}
\bar \tna &=&\Bigl \langle \sum_i \delta(\vec x - \vec x_i) 
	   \psia (\eta(m_i)) \Bigr\rangle \nn \\
   &=& \int_0^\infty dm \Bigl\langle \sum_i \delta(m-m_i)
	    \delta(\vec x - \vec x_i) \psia(\eta(m)) \Bigr\rangle 
	     \nn \\
 &=&  \int_0^\infty dm 
	   \Bigl \langle \sum_i \delta(m-m_i) \delta(\vec x - \vec x_i) 
	   \Bigr \rangle \langle \psia(\eta(m))\rangle \nn.
\end{eqnarray}
Notice that is only because we are assuming that $\eta$
is independent of cosmology that we can 
take $\psia( \eta (m))$ out of the ensemble average above, and consider its 
average value over the halo occupation distribution alone.\footnote{This
is a crucial and strong assumption.  In particular, any dependence of the
chosen mass tracer on the large scale environment would bias our results.}
We now define the quantity
\begin{equation}
n(m,\vec x) \equiv  \sum_i \delta(m-m_i) \delta(\vec x - \vec x_i), 
\label{mf}
\end{equation}
which represents the true halo density field, the spatial average of which
is called the halo mass function $\bar n(m)$.   We obtain then
that the number density of groups in a bin $\vec a$ is
\begin{eqnarray}
\bar \tna
& = & \int_0^{\infty} dm \,\langle n(m,\vec x) \rangle \,
    \langle \psia (\eta(m)) \rangle \nn \\
     & = &  \int_0^{\infty} dm\, \bar n(m) \avgpsia. 
\label{tna}
\end{eqnarray}
where $\avgpsia$ is meant to be the average 
value of $\psia(\eta(m))$ over the probability distribution $P(\eta|m)$.

\subsection{The Cluster-Cluster Correlation Function and Power Spectrum}

Following an argument  similar to the one above we may obtain an expression
for the cluster-cluster correlation function $\tilde\xi_{\vec a,\vec a'} (r)$ 
between objects in bins $\vec a$ and $\vec a'$ 
(here $r=|\vec x - \vec x'|$):
\begin{eqnarray}
\xi_{\vec a,\vec a'}^t (r) & = & \langle 
      (\frac{\tna(\vec x)}{\bar \tna}-1)
      (\frac{\tnap (\vec x')}{\bar \tnap}-1)
      \rangle \nn \\
& = & -1+ \int_0^{\infty} dm\, dm' 
     \avgpsia \avgpsiap
 \,\frac{\langle n(m,\vec x)  n(m',\vec x')\rangle}{\bar\tna\bar\tnap} \nn \\
& = & -1+ \int_0^{\infty} dm\, dm' 
       \avgpsia \avgpsiap \frac{\bar n(m) \bar n(m')}{\bar \tna \bar \tnap}
      \bigl(1+\xi_{hh}(r|m,m')\bigr) \nn \\
& = & \int dm\, dm' \xi_{hh}(r|m,m')
      \avgpsia \avgpsiap  \frac{\bar n(m) \bar n(m')}{\bar \tna \bar \tnap} 
	\nn 
\end{eqnarray}
where $\xi_{hh}(r|m,m')$ is the halo correlation function 
between halos of different masses $m$ and $m'$ separated by distance $r$.
The upper script $^t$ on $\txia$ serves here again to remind us that this is 
the true cluster correlation function. The observed correlation 
function may differ from $\txia$ due to systematics effects in the
observations, to be included later.
From our discussion concerning the average cluster density, we already 
know how to handle $\bar \psia(m)$ and  $\bar n(m)$.  The only term 
in the above expression which is new is $\xi_{hh}$.

The halo model provides a prescription to relate $\delta_{h}$, the 
halo overdensity, to the matter overdensity. This same prescription
relates the halo correlation function to the linear correlation
function. To first order in perturbation 
theory, one obtains (see e.g. \cite{CooraySheth}, \cite{ShethTormen})
$\delta_h= \bar b(m)\delta_m^{\rm lin}$ where
\begin{eqnarray}
 \bar b(m) & = & 1-\frac{\partial \ln \bar n(m)}{\partial \delta_{sc}} \nn \\
& = & 1+\frac{q\nu -1}{\delta_{sc}} 
			  + \frac{2p}{\delta_{sc}(1+(q\nu)^p)} 
\label{eq:bias}\end{eqnarray}
is the linear bias \cite{ShethTormen}\footnote{We have chosen to denote this
quantity $\bar b$ since the expression depends on the average halo mass 
function $\bar n(m)$ rather than $n(m,\vec x)$.}. Here $\delta_{sc}$ is
the critical overdensity needed for collapse, $\delta_{sc}=1.686$; 
$q=0.75$ and $p=0.3$ come from fitting to $N$-body simulations; and $\nu\equiv
\delta_{sc}^2/\sigma_{\rm lin}^2(m)$, with $\sigma_{\rm lin}$ being the tophat
filtered rms fluctuations in the linear density field. The radius $R$ of 
the top-hat filter 
used in defining $\sigma(m)$ is obtained by demanding that a sphere of
radius $R$ encompasses a total mass $m$, i.e. 
$\Omega_m\rho_c 4\pi R^3/3 = m$, with $\rho_c$ the critical density
of the universe. 

In this limit, the halo-halo correlation function becomes 
 $\xi_{hh}(r|m,m') = \bar b(m) \bar b(m')\xi_{\rm lin}(r)$
and thus the cluster-cluster correlation function simplifies to:
\begin{equation}
\xi_{\vec a,\vec a'} (r) = \bar\tba\bar\tbap \xi_{\rm lin}(r)
\end{equation}
where
\begin{equation}
\bar \tba \equiv \int_0^\infty dm \frac{\bar n(m)}{\bar\tna} \bar b(m) 
	       \bar \psia (m).
\label{bias}
\end{equation}
Not surprisingly, we see that the cluster-cluster correlation function
traces the underlying mass correlation function.
It is worth pointing out that observationally the power spectra of galaxies 
and clusters are seen to have the same shape but different amplitude on
scales  $k^{-1} \sim 10$ Mpc or larger, so that the simple scale-independent 
linear bias is indeed feasible on large scales \cite{Tadros} (in fact, it
is necessarily so for guassian fields \cite{ScherrerWeinberg}).

\section{Intrinsic Errors Estimate}

We discuss now three intrinsic  uncertainties associated with 
our observables: Poisson errors, sample variance, and 
uncertainties due to intrinsic scatter in the mass-tracer to mass
scaling relation. Keeping an eye on our choice of model survey in
\S V,
we assume a volume limited sample such that all objects of interest 
(i.e. all halos with $\eta$ in the range of interest)
are detected.  Further, we assume no contamination of the sample, and
a perfect instrument so that the observed value 
$\eta^{obs}$ of the mass tracer always matches the true value $\eta^{true}$.
Experimental bias and scatter are treated in \S IV.

\subsection{Poisson Uncertainties and Sample Variance}

The first type of uncertainty is the Poisson error in
the number of clusters found.  Assuming no intrinsic clustering, the
variance in the density of clusters is simply given by $\bar\tna/\bar V$.
This contribution to the correlation matrix is therefore
\begin{equation}
\tC^{nn}\eval_{Poisson} \equiv \langle \big( n_{\vec a} - 
\bar\tna \big) \big(
n_{\vec a'} - \bar n_{\vec a'}^t\big)\rangle^{\rm Poisson} 
	= \da\frac{\bar\tna}{\bar V}.
\end{equation}
Note the use of the symbol $\tC$ for the correlation matrix.  We include
a $\sim$ above the $C$ to identify it as the ``intrinsic" correlation 
matrix, meaning that no observational effects have been taken into 
account. There is also a Poisson term in the bias arising from
the estimation of the clustering properties with a finite number of galaxies
(namely the secon term in equation 34) which is treated later on to 
properly account for contamination and completeness.

In addition to this,
Hu and Kravtsov showed that sample variance becomes increasingly important
as we probe lower and lower mass scales \cite{HuKravtsov} in the halo mass
function.  We rederive here
the result for sample variance found in Hu and Kravtsov, to use it 
as a reference for deriving the sample variance errors involving bias.

Assuming we have averaged over $P(\eta|m)$ (uncertainties associated with
this probability are derived below),
the sample variance contribution to the density-density correlation matrix 
is given by
\begin{eqnarray}
\tC^{nn}\eval_{sample} =  \int d^3xd^3x' W(\vec x) W(\vec x') 
	   \int dmdm'\avgpsia \avgpsiap 
        \bigl\{ \langle n(m,\vec x)n(m',\vec x') 
	     \rangle -  \bar n(m) \bar n(m')  \bigr\} \nn
\end{eqnarray}
where $W(\vec x)$ is the survey's window function.  Note we do 
not include terms proportional to simultaneous deviations from the mean 
due to sample variance and the variance of $\psia$ due to $P(\eta|m)$ as 
these would yield only small corrections.

We approximate $W(\vec x)$ above 
to be a spherical top-hat function encompassing a volume equal to that of the
survey volume.
The matrix element can now be easily computed by replacing $n(m,\vec x)$ in the
above expressions with $\bar n(m)+\delta n(m,\vec x)$ where 
$\delta n = \bar b(m)\bar n(m)\delta$.
Since $\langle \delta \rangle =0$ and 
$\langle \delta \delta \rangle=\xi(\vec x - \vec x')$, we obtain then
\begin{eqnarray}
\tC^{nn}\eval_{sample} 
	& = &  \bar \tna \bar \tnap \bar \tba \bar \tbap
		  \int d^3k P(k) |W(k)|^2 \nn \\
       & = &  \bar \tna \bar \tnap \bar \tba \bar \tbap
	      \sigma^2(R_V).
\end{eqnarray} 
where $R_V$ is given
by $4\pi R_V^3/3 =V$. This is the 
final result that we were looking for, namely an expression for the 
covariance matrix between the number of objects found in each bin due
to sample variance.  The total density-density correlation matrix is obtained 
then by adding the sample variance matrix and the Poisson noise.

Let us now turn towards matrix elements involving bias.
As before, the sample variance contribution is 
\begin{eqnarray}
\tC^{nb}\eval_{sample} = \int d^3xd^3x' W(\vec x) W(\vec x')  \int dmdm'
  \avgpsia \avgpsiap
  \{ \langle n(m,\vec x)b(m',\vec x') \rangle - 
    \bar n(m) \bar b(m')   \}  \nn
\end{eqnarray}
Again, we replace $n(m,\vec x)$ by $\bar n(m)+\delta n(m,\vec x)$ and 
$b(m,\vec x')$ by $\bar b(m)+\delta b(m,\vec x)$.
To get an expression for $\delta b$, we generalize Eq.~(\ref{eq:bias}) to
unbarred $b$ and $n$, so that 
\begin{eqnarray}
\bar b+\delta b & = & 1-\frac{\partial}{\partial \delta_{sc}}
					    \big( \ln (\bar n(1+\bar b\delta)) \big) \\
	   & = & \bar b - \delta \frac{\partial \bar b}{\partial \delta_{sc}}
\end{eqnarray}
where the second equality holds to leading order in $\delta$.  Hence
\begin{equation}
\delta b = - \delta \frac{\partial \bar b}{\partial \delta_{sc}}.
\end{equation}

With this result, the sample variance contribution to $\tC^{nb}$ 
may be written as
\begin{equation}
\tC^{nb}\eval_{sample} = \bar\tna\bar\tba\bar\tbap\bar\tcap 
	\sigma^2(R_V)
\end{equation}
where 
\begin{equation}
\bar\tca = \int dm \frac{\bar n(m)\bar b(m)}{\bar n_{\vec a}\bar b_{\vec a}}
	 \bar \psi_{\vec a}(m)  (-\frac{\partial \bar b}{\partial \delta_{sc}})
\end{equation}

In exactly the same way we obtain the bias-bias terms of the 
correlation matrix, which are given by 
\begin{equation}
\tC^{bb}\eval_{sample}=\bar\tba\bar\tbap\bar\tca\bar\tcap
	\sigma^2(R_V).
\label{bbSV}
\end{equation}

\subsection{Mass Tracer Dispersion Errors}

By mass tracer dispersion errors we mean the uncertainties in the
density and bias of objects in a given bin $\vec a$ due to the fact 
that $\eta$ is not uniquely determined by $m$, i.e. uncertainties due to the
probability distribution $P(\eta|m)$.  
Note these are somewhat analogous to Poisson uncertainties in the number 
of galaxies in a spatial bin.  
Not surprisingly, then, the uncertainties take on a form
$\delta n \sim n/\bar V$ where $\bar V$ is the volume of the survey.

Consider then a realization of 
the halo model.  Given the survey volume $\bar V$, the number density
of objects in bin $\vec a$ in one realization may be written as
$$
\tna=\frac{1}{\bar V}\sum_i \nu_i\psi_{\vec a}(\eta(m_i))
$$
where $\nu_i$ is the number of halos in the mass bin $m_i$.  The mass 
binning must be chosen small enough so that 
at most one halo is found in any given mass bin.  This ensures $\eta_i$
is a random variable for each halo (i.e. that $\eta_i$ and $\eta_j$ are
uncorrelated for any two halos $i,j$). We have then 
\begin{eqnarray}
\overline{\tna\tnap} & = & \frac{1}{\bar V^2}
   \sum_{i}\Big\{\sum_{j\neq i} \nu_i\nu_j 
   \langle\psia(\eta(m_i))\psiap(\eta(m_j)) \rangle + \mbox{ } \nu_i^2 
    \langle \psia(\eta(m_i))\psiap(\eta(m_i)) \rangle \Bigr\} \nn \\
& = & \frac{1}{\bar V^2}
      \sum_i\Bigl\{\sum_{j\neq i} \nu_i\nu_j\langle \psi|m_i\rangle 
	\langle\psiap|m_j\rangle
      + \mbox{ }\nu_i \langle\psia(\eta(m_i))\psiap(\eta(m_i))\rangle \Bigr\}
\label{eq:nbarone}\end{eqnarray}
We have used above that $\eta(m_i)$ and $\eta(m_j)$ are not correlated, 
and that $\nu_i^2 = \nu_i$ (since $\nu_i=0,1$).  We may write a similar
expression for $\bar\tna\bar\tnap$,
\begin{equation}
\bar\tna\bar\tnap  =  \frac{1}{\bar V^2}
      \sum_i\Bigl\{\sum_{j\neq i} \nu_i\nu_j\langle \psia|m_i\rangle
	\langle\psiap m_j\rangle \nn \\
	 +  \nu_i \langle \psia|m_i\rangle \langle\psiap |m_i\rangle \Bigr\} 
\label{eq:nbartwo}
\end{equation}
so upon subtracting Eq.~(\ref{eq:nbartwo}) from Eq.~(\ref{eq:nbarone}), we find
$$
\tC^{nn}\eval_{\mbox{\it mass tracer}} 
	= \frac{1}{\bar V^2} \sum_i \nu_i \C^{\psi\psi}(m_i)
$$
where 
\begin{equation}
\C^{\psi\psi} (m) = \langle \psia(\eta(m))\psiap(\eta(m))\rangle -
\avgpsia\langle\psi_{\vec a'}|m\rangle
\label{varpsi}
\end{equation}

But note that $\nu_i/\bar V$ is just the average number density of halos in 
a bin of mass $m_i$.  Averaging over many realizations, we have
$$
\Bigl\langle \frac{\nu_i}{\bar V} \Bigr\rangle = \bar n(m_i)\Delta m_i
$$
and hence
$$
\tC^{nn}\eval_{\mbox{\it mass tracer}} 
	= \frac{1}{\bar V} \sum_i \Delta m_i \bar n(m_i)\C^{\psi\psi}(m_i).
$$

In the continuum limit, we obtain our final answer,
\begin{equation}
\tC^{nn}\eval_{\mbox{\it mass tracer}} 
	= \frac{1}{\bar V}\int dm \bar n(m) \C^{\psi\psi}(m)
\end{equation}
where $\C^{\psi\psi}$ is given by equation (\ref{varpsi}).  If the
binning function satisfies
$\psia(\eta)\psiap(\eta) = \da \psia(\eta)$ (e.g. non-overlapping 
top-hat bins)
equation (\ref{varpsi}) simplifies to
\begin{equation}
\C^{\psi\psi}(m) = \da\avgpsia-\avgpsia\langle\psiap|m\rangle.
\label{eq:cpsi}\end{equation}
Our expression for the contribution to the correlation matrix from
$P(\eta|m)$ makes sense. If $P(\eta|m)$ is very narrow, then a given
mass $m$ will always get assigned to one and only bin $\vec a$. In that
case, $\avgpsia=1$, and Eq.~(\ref{eq:cpsi}) shows that the correlation 
matrix vanishes. That is, if the mass tracer is perfect, then there is no 
uncertainty associated with it. In the more realistic case, a mass $m$ will 
sometimes be assigned to more than one bin.
Then, the second term in Eq.~(\ref{eq:cpsi}) would be non-zero even
if $\vec a\ne \vec a'$  This would lead then to an anti-correlation
between these two bins. Because of the leakage into adjoining bins, 
$\avgpsia<1$
and the diagonal term in $C^{\psi\psi}$ also becomes nonzero.

What about density-bias terms?  Going back to equation (\ref{bias}),
 we can write the bias as
$$
\tba = \frac{1}{\bar\tna \bar V}\sum_i\nu_i b(m_i)\psia(\eta(m_i))
$$
A procedure analogous to the one before yields
$$
\tC^{nb} = \frac{1}{\bar\tnap \bar V^2} \sum_i \nu_i b(m_i) \C^{\psi\psi}(m_i).
$$
Converting this into an integral, we get
\begin{equation}
\tC^{nb} = \frac{1}{\bar\tna \bar V}
	   \int dm \bar n(m) \bar b(m) \C^{\psi\psi}(m)
\end{equation}
Incidentally, note that the correlation matrix is still symmetric despite
the appearance of a factor $1/\bar\tnap$.  The 
matrix element across the diagonal is $\tilde C^{bn}_{\vec a',\vec a}$,
which also contains the same $1/\bar\tnap$ factor.  All other terms
in the expression are clearly symmetric.

Finally, performing the same analysis yields the bias-bias contribution,
given by
\begin{equation}
\tC^{bb} = \frac{1}{\bar V}\frac{1}{\bar\tna\bar\tnap}
	   \int dm \bar n(m)\bar b(m)^2 \C^{\psi\psi}(m).
\end{equation}

\section{Observation Related Uncertainties and Systematics}

We have derived above expressions for the density and bias of halos
binned according to an unspecified mass tracer $\eta$,
along with the related intrinsic uncertainties assuming a volume limited
survey.  We have, however, been assuming
no contamination, $100\%
$ rate detection (completeness), 
and the ability to observe
$\eta$ precisely.  We wish to incorporate into our formalism
uncertainties and systematic effects arising from observation.  In 
particular, we assume observational effects may
be characterized by the following information:

\begin{itemize}
\item The average detection rate (completeness) $\bar r_{\vec a}$ and 
      its variance $C_{\vec a,\vec a'}^{rr}$. $\bar r_{\vec a}$ is defined as 
      the fraction of objects in bin $\vec a$ which are identified.  
\item The average false detection rate $\bar f_{\vec a}$ and its variance
      $C_{\vec a,\vec a'}^{ff}$. $\bar f_{\vec a}$ is defined as the number of 
      spurious objects per unit volume in bin $\vec a$.  
\item The probability $q(\eta|\eta^t)$ that a halo with observed 
      $\eta$ has a true value $\eta^t$.
\end{itemize}	 

Note we made the simplifying assumption that both 
$\ra$ and $\fa$ are position independent, though it is straightforward
(albeit cumbersome)
to extend the formalism to include position dependent detection and 
contamination rates. 
 
We begin by incorporating detection and contamination rates
into the formalism.   Further, since the true observable is not 
the density of objects but rather the number of objects in a given bin, 
we modify our formalism appropriately.  Given our assumptions, the number of 
objects identified in a volume limited survey takes the form
\begin{equation}
\Na = (r_{\vec a}\tna +f_{\vec a})V
\label{obsNa}
\end{equation}
where $\tna$ is the average number of objects in bin $\vec a$.
The corresponding correlation matrix is then\footnote{Throughout, we are 
using the fact that 
$$
\alpha_i=a_ib_i+c_i \Rightarrow 
C_{ij}^{\alpha\alpha} = \bar a_i\bar a_j C_{ij}^{bb}
		        + \bar b_i \bar b_j C_{ij}^{aa} + C_{ij}^{cc}.
$$
Terms involving products of correlation matrices are being ignored as
second order terms.  For the expression above, $a_i,b_i$, and $c_i$
are all uncorrelated with each other and possess different probability
distributions (which is the case of interest here).}
\begin{eqnarray}
\C^{NN} & = & \C^{rr}\bar \tna \bar \tnap \bar V^2 +
	\bar\ra \bar\rap \bar V^2 \tC^{nn} + \C^{ff}\bar V^2 
     +  \frac{\bar \Na}{\bar V}\frac{\bar \Nap}{\bar V}(\Delta V)^2 \nn \\
& =  &  \frac{\bar \Na -\bar \fa\bar V}{\bar\ra}
    \frac{\bar \Nap -\bar \fap\bar V}{\bar\rap} \C^{rr}
    + \bar V^2\C^{ff} + \bar\ra\bar\rap\bar V^2\tC^{nn} 
    + \bar\Na\bar\Nap \frac{(\Delta V)^2}{\bar V^2}
\end{eqnarray}
where $(\Delta V)^2$ is the variance in the volume arising from uncertainties
in the measured redshifts of the clusters.  The result is just what
one would expect: the uncertainties for a perfect algorithm $\tC^{nn}$
are scaled by the detection factors.  In addition, one adds the uncertainties
due to imperfect knowledge of the detection and false detection
rates.  Finally one needs to include the contribution from uncertainties
in the sampled volume arising from redshift uncertainties. 
The expression above simplifies
when the detection and contamination rates of different bins are 
uncorrelated, in which case one obtains 

\begin{equation}
\C^{NN} = \da\Bigl\{(\bar\Na-\bar\fa\bar V)^2
	     \frac{(\Delta \ra)^2}{\bar\ra^2} 
	     + (\bar V \Delta \fa)^2\Bigr\} + \bar\ra\bar\rap\bar V^2\tC^{nn} 
    + \bar\Na\bar\Nap \frac{(\Delta V)^2}{\bar V^2}.
\end{equation}

We wish to incorporate now systematic effects
and/or uncertainties in the assigned cluster richness 
arising from the observations.
In particular, we wish to include the fact that binning of data is done 
in terms of the \it observed \rm value of $\eta$ rather than the 
true value $\eta^t$.  We view the observations as providing
a random mapping $\eta(\eta^t)$ where a value $\eta$ has a probability
$q(\eta|\eta^t)$ of occurring.  The probability $q(\eta|\eta^t)$ is 
assumed to be known from experimental calibrations. e.g. if we consider
X-ray temperature measuremetns for a cluster, $q(T|T^t)$ describes how a 
series of measuremnts $T_i$ of a cluster with temperature $T^t$ 
is distributed.

Consider then the number of objects in a bin $\vec a$.  All we need
to do then is replace $\psia(\eta^t)$ by $\psi(\eta)$ where $\eta$
is the observed value.  The relevant probability for computing
$\bar \psia (m) = \langle \psia(\eta) |m \rangle$ is not 
$P(\eta^t|m)$ but $\tilde P(\eta|m)$, the probability of observing a
value $\eta$ given a halo mass $m$.  This is given by
\begin{equation}
\tilde P(\eta|m) = \sum_{\eta^t} q(\eta|\eta^t)P(\eta^t|m).
\label{probconv}
\end{equation}
By replacing $\eta$ by $\eta^t$
and $P(\eta^t|m)$ by $\tilde P(\eta|m)$ in formulae, 
we obtain expressions for the
number density and bias of halos binned according to the observed value
$\eta$.  Any systematic effects and/or uncertainties introduced by the
observations will automatically be taken into account in the
convolution of $q(\eta|\eta^t)$ and $P(\eta^t|m)$. Note though that
an application of our formalism requires an understanding of the 
probability $q(\eta|\eta^t)$, presumably characterized in the
experiment's calibration.

Let us now turn our attention to bias.  The density 
of  detected groups in bin $\vec a$, which we shall denote 
$\na$ (note the superscript $^t$ is missing), is given by
$$
\na(\vec x) = \ra \tna(\vec x) + \fa
$$
where we are assuming a detection rate $\ra$ and a false detection rate 
$\fa$, known from calibration to have values $\bar\ra$ and $\bar\fa$
and variances $\C^{rr}$ and $\C^{ff}$.\footnote{Note 
we are making use of our simplifying assumption
that $\ra$ and $\fa$ do not depend on position.  Otherwise, there would
be systematic corrections to equation (\ref{obscorr})
due to the correlation functions of $\ra$
and $\fa$. These effects are easily incorporated, but we choose not to 
in the spirit of simplicity.}
With these assumptions, the observed correlation function is given by
\begin{equation}
\bar\na\bar\nap\xia^{nn} = \bar\ra\bar\rap\bar\tna\bar\tnap\xia^{t,nn}.  
\label{obscorr}
\end{equation}
Replacing $\xia^{t,nn} = \tba\tbap\xi_{\rm lin}$, and dividing
through by $\bar\na\bar\nap$ we obtain
\begin{equation}
\xia^{nn} = \ba \bap \xi_{\rm lin}
\end{equation}
where
\begin{eqnarray}
\ba & = & \frac{\bar\tna\bar V}{\bar\Na}\bar\ra \tba \nn \\
    & = & \Bigl(\frac{\bar\Na-\bar\fa\bar V}{\bar\Na}\Bigr)
	\tba. \label{obsba}
\end{eqnarray}
The expectation value of the observed bias is therefore simply
\begin{equation}
\bar\ba = \Bigl(\frac{\bar\Na-\bar\fa\bar V}{\bar\Na}\Bigr) \bar\tba.
\end{equation}
The observed bias is thus expected to be lower than the true bias 
due to the dilution of true clusters with false detections.  The clustering
signal, however, does not depend on $\ra$.
This is not surprising, as $\ra$ changes only how many clusters we 
observe, but not their clustering properties.  Of course, a lower $\ra$
will lead to more noisy estimations of the bias (see equation 
\ref{biaserrest}).

Now that we have our expression for bias, we may compute the 
corresponding uncertainties.  Using equations (\ref{obsNa}) and
(\ref{obsba}), we obtain
\begin{equation}
\C^{Nb}  = \Bigl(\frac{\bar\Nap-\bar\fap\bar V}{\bar\Nap}\Bigr)
	  (\bar\ra\bar V\tC^{nb}).
\end{equation}

We can understand the expression above qualitatively.    
The factor $\bar\ra\bar V$ in front
of $\tC^{nb}$ scales the correlation matrix to the total number
of real clusters as opposed to number density. The prefactor
corresponds to the scaling from the true bias to the observed bias.

Finally, this same type of analysis gives us the bias-bias 
terms of the observable's correlation matrix, resulting in
\begin{equation}
\C^{bb}  =  \Bigl(\frac{\bar\Na-\bar\fa\bar V}{\bar\Na}\Bigr)
	  \Bigl(\frac{\bar\Nap-\bar\fap\bar V}{\bar\Nap}\Bigr)
	  \tC^{bb}.
\end{equation}

In addition to the errors above, there is a contribution
coming from actually attempting to estimate bias from data.
For instance, assuming the bias is measured using the power spectrum
involves Fourier space discretization.  This,
in turn, brings its own set of sample variance errors, as well as Poisson
errors in the number of objects
found in a Fourier pixel.  The corresponding
errors in the power spectrum estimation are given by 
(see e.g.\cite{Scott,Tegmarketal,Bernardeauetal}):
\begin{equation}
\Bigl( \frac{\Delta \ba (k)}{\ba} \Bigr)^2 \approx 
       \frac{2}{VV_k}\Bigl(1+\frac{1}{\bar \na  (\bar\ba)^2 P(k)}\Bigr)
\label{biaserrest}
\end{equation}
where $V$ is the volume of the survey, and $V_k$ is the volume of the 
corresponding $k$-shell in Fourier space.
i.e. $V_k \approx 4\pi k^2 \Delta k$ where $\Delta k$ is the minimum spacing
between $k$ modes.  
Note the error estimate above $\Delta \ba (k)$ depends on
the wavenumber $k$ used to estimate $\ba$.  One may average the
estimated $b(k)$ over a large $k$ range to obtain smaller errors.
This diagonal contribution to the bias-bias correlation
matrix must be added to that in equation (\ref{bbSV})
since it represents uncertainties in the
experimental estimation of the bias parameter.

\section{Application: Group and Cluster Statistics in a Volume Limited 
Galaxy Survey}

We now apply our formalism to a hypothetical cluster catalogue
obtained from a large volume limited galaxy survey.  We are interested
in particular in what kind of information we can extract
from such a catalogue.  We present below our assumptions as to how the
hypothetical catalogue is built, followed by how the halo model formalism 
is applied in this particular case.  We also present
the fiducial model used in the next section to
derive the type of
constraints one can expect for these type of surveys.

\subsection{Assumptions on the Cluster Catalogue}

\begin{figure}[t]
\sfig{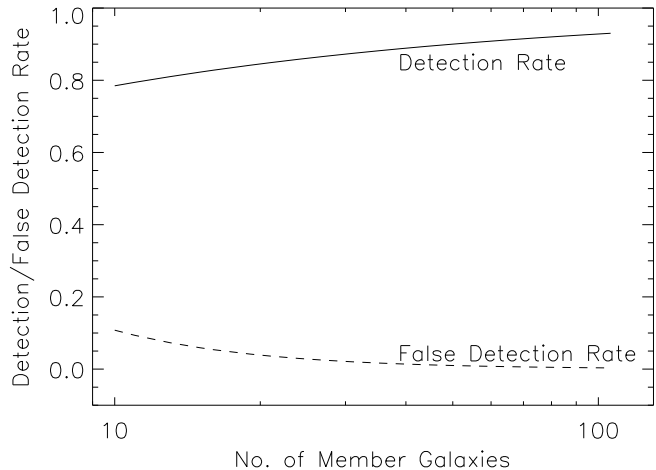}{0.55\columnwidth}
\caption{\footnotesize The assumed detection and false detection rates 
are shown here with the solid and dashed line respectively.
Note the latter is defined here in terms of the
percentage of identified cluster that are false detections.}
\label{decrates}
\end{figure} 

Given a galaxy sample, one may attempt to identify groups and 
clusters of galaxies within it.  While the notions of clusters and 
cluster richness are intuitive, ultimately one needs precise
definitions to obtain a well defined sample.  This task
is achieved via automated cluster finding algorithms.
At present, there exist a large number of said algorithms, 
e.g. maxBCG \cite{Annisetal}, Hybrid Match Filter (HMF) \cite{Kimetal},
Cut and Enhance method (CE) \cite{Gotoetal}, Vornoi Tessellation Technique
(VTT) \cite{Kimetal}, C4 algorythm (\cite{C4},\cite{Nichol}), and
Friends-of-Friends (FoF) \cite{HuchraGeller},
all of which simultaneously identify clusters (based on a specific
set of criteria) and assign a richness measure
to the identified clusters.  The richness measures are typically 
the number of galaxies assigned to the cluster, or some measure of the
cluster's total optical luminosity.
If we identify clusters with massive halos, the richness measure provides
then an observable which serves as a mass indicator.  We can thus
use our formalism to describe statistical
properties of cluster catalogues by using the richness as the mass tracer
$\eta$. 

In this paper, we assume a cluster finding algorithm having a specified
set of criteria to determine when a galaxy belongs (is assigned to) 
a cluster.  The number of galaxies $N$ assigned to a cluster will be our 
richness measure.  Examples of this type of algorithm are maxBCG, C4,
Friends-of-Friends, Vornoi Tessellation Technique\footnote{The
Vornoi Tessellation Technique as given in Kim et al. \cite{Kimetal}
assigns richness using the match filter method.  Nevertheless, one
could imagine using the number of galaxies $N$ that the VTT technique
assigns to the cluster as a richness measure.}, and the cut and
enhance method.  For our hypothetical catalogue, we will not worry
about what the exact criteria for cluster membership is for a galaxy: we 
simply need to assume such a criteria exists.\footnote{We note though
that the membership criterion will in general affect the expected 
mass-richness relation.e.g. the number of member galaxies will clearly 
depend on the radius used to determine membership.}

Within this framework, then, the mass tracer is the number of galaxies
in a halo $N^t$ while the observed richness measure is $N$, the number
of galaxies assigned to a cluster by the cluster finding algorithm.
$q(N|N^t)$ is the probability that the cluster finding algorithm
will assign $N$ galaxies to a halo containing $N^t$ galaxies.  
Likewise, $\bar\ra$
and $\bar\fa$ are the detection and false identification rates for
the cluster finding algorithm.\footnote{Note we are not considering
q to give rise to false detections or imperfect detection rates. If a 
cluster is identified,but its richness is mislabelled, that effect is
encoded in $q(N|N^t)$.  On the other hand, a cluster that is broken up
into two smaller clusters, or merged with another cluster to produce
a larger one cannot be considered as a mislabelled cluster.  In particular,
these last two effects would greatly alter the richness and ruin the 
one-to-one and
onto nature of the mapping between $N$ and $N^t$ that we have been assuming.
The inclusion of detection rates and false detections here serves to naively
account for these effects.}

We make the further assumption that
$\bar\ra$ and $\bar\fa$ are uncorrelated between various bins and
with each other.

It is worth pointing out that at least some cluster finding algorithms
(e.g. match filter algorithms) have a detection rate that is dependent
on the galaxy background (see e.g. Kim et al. \cite{Kimetal}).  We 
will ignore this effect here.  To include it, one could imagine
the detection rate having the form $\ra=\bar\ra(1+\gamma\delta_g)$
where $\delta_g$ is the galaxy density contrast and $\gamma$
is a constant.  Notice $\gamma$ is a measure of whether $\ra$ is strongly
dependent on the background density or not.  
Using $\delta_g \approx \delta$ (since galaxies
are unbiased tracers of mass), we could replace the above expression 
in equation 22 and rederive the corresponding uncertainties as in
the previous section.  This would add terms proportional to $\gamma$,
which may be neglected in the limit that $\gamma$ goes to zero.  
We do not expect this effect
to have major consequences in our results.

Let us then specify the characteristics of our hypothetical cluster finding
algorithm.  We will assume that the non-detection rate (i.e. $1-\bar\ra$),
the false detection rates, and their errors, are all power laws
as a function of the number of galaxies in the cluster.
Thus, e.g., the detection rate is taken to have the form
\begin{equation}
\bar r_N = 1-(N/N_0)^{-\gamma}
\end{equation}
where $N_0$ and $\gamma$ are constants. The normalization is
specified by indicating their corresponding values for clusters
with $5$ galaxies and clusters with $50$ galaxies, shown below, as well
as the corresponding values $N_0$ and $\gamma$.
Figure \ref{decrates} plots the corresponding rates.

\begin{center}
\begin{tabular}{|c|c|c|c|c|}
\hline
Rate/Error        & N=5		& N=50    & N$_0$    & $\gamma$ \\ \hline
$\bar r_N$	  & $70\%$	& $90\%$  & 0.400    & 0.477 \\ \hline
$\Delta \bar r_N$ & $10\%$	& $1\%$   & 0.500    & 1.000 \\ \hline
$\bar f_N$	  & $30\%$	& $1\%$   & 2.213    & 1.477 \\ \hline
$\Delta \bar f_N$ & $10\%$	& $1\%$   & 0.500    & 1.000 \\ \hline
\end{tabular}
\end{center}

Note that the false detection rate 
corresponds to the percentage of detected clusters in the fiducial
model which are false detections.  Thus, e.g. we are assuming that
$30\%
$ of all detected clusters with 5 member galaxies are false detections.
The correct values for the different type of algorithms vary, but we
believe that the numbers above should provide a fair picture of the
capabilities of cluster finding algorithms at low redshifts.
Details on particular algorithms may be found
in the references (see e.g. \cite{Annisetal}, \cite{Kimetal}, \cite{Gotoetal}, 
\cite{Kimetal}, \cite{C4}, \cite{Nichol}, \cite{HuchraGeller}).  

Also note that when we apply our formalism we will need to assume that
the data is binned into various richness classes, i.e. into bins of 
clusters containing $N$ galaxies where $N_{max}>N\geq N_{min}$.  We
will use the same detection rate for all objects in a given bin, the 
detection rate being defined as the average rates for clusters with 
richness $N_{min}$ and $N_{max}$.

Finally, we need to specify the probability that the algorithm assigns
$N$ galaxies to a cluster given that its parent halo has $N^t$
galaxies (i.e. what we had called $q(\eta|\eta^t)$ earlier).  
It is difficult to find within the literature expressions 
for this probability.  Here, we assume that the number of galaxies assigned
to a halo takes the form $N=N^t+\delta N$ where $\delta N$ is a random
variable with an exponential distribution.\footnote{There is no
reason to choose the distribution we used other than
it is simple.  We expect, however, that this distribution is at least
qualitatively correct.} In other words, we take 
$q(N|N^t)=P(\delta N)$ with $P$ given by
\begin{equation}
P(\delta N|N^t) = A\exp(-a(N^t)|\delta N|).
\label{deltaNdist}
\end{equation}
The parameters $A,a$ in the above expression are determined by 
the condition that the probabilities add to one, and by requiring
that the expectation value of $|\delta N|$ be $10\%$ of 
$N^t$.\footnote{Again, the average value of $|\delta N|$ is arbitrarily
chosen but we expect it to be representative.}
Note this distribution is wider than a Gaussian.

It is often the case that cluster finding algorithms systematically 
underestimate the number of galaxies of a cluster.  This effect is
easily accommodated by correcting our expression for $N$ to 
$N=f(N^t)+\delta N$ where $f(N^t)$ is the average number of galaxies assigned
to clusters with $N^t$ galaxies.  Since the only effect this brings
about is a rescaling of the axis, we do not expect our conclusions
to be changed due to possible biases in the galaxy assignments of 
cluster finding algorithms (provided, of course, that they are appropriately
calibrated).

As a closing note, we would like to emphasize that 
all results presented here depend on the ability to accurately 
calibrate cluster finding algorithms. In particular, recall  the
parameter $a(N^t)$ in equation \ref{deltaNdist} is determined by demanding
the expecation value of $|\delta N|$ to satisfy 
$\langle|\delta N|\rangle = cN^t$ where the value $c=10\%$ was arbitrarily
chosen. 
In general,
marginalization over the $c$ parameter as determined from calibrations 
will also be necessary, leading to 
a degration of the confidence regions presented here.

\subsection{The Halo Occupation Distribution}

In order to apply our formalism, we need an expression for $P(N^t|m)$,
the probability for a halo of mass $m$ to have $N^t$ galaxies in it.
This probability is known as the Halo Occupation Distribution,
or HOD.
In accordance with the results of  Kravtsov et al. \cite{AndreyAndreas}, 
we assume all halos have 
a central galaxy, while any other galaxies found in the halo (referred to as
satellite galaxies) are 
Poisson distributed with an average number
\begin{equation}
\langle N_{sat}|m \rangle = (m/M_1)^\alpha.
\label{avgnumgal}
\end{equation}
Here $M_1$ is the normalization parameter.  It represents the 
mass of halos at which one expects, on average, to find $1$ satellite 
galaxy.  Prior galaxy formation simulations in which the distinction between 
host and satellite galaxies was not made agree with the results 
from Kravtsov et al.
in that the number of galaxies in halos with large occupancy numbers is Poisson
distributed with the average number increasing as a power law 
\cite{Andreas}.
Furthermore, the power law assumption for $\avgN$ has
been used to model the galaxy correlation function for both the 2dF survey
and the SDSS survey with very good agreement (see e.g. 
Magliocchetti and Porciani \cite{MagliocchettiPorciani} for the 2dF and 
Zehavi et al. \cite{Zehavietal} for the SDSS.)  One may also wonder whether
there is evidence that the probability $P(N^t)$ of a halo does depend indeed 
exclusively on the mass. Once
again, simulations seem to indicate that this is indeed the case
\cite{Andreas}.

\subsection{Fiducial Model}

\subsubsection{Cosmology}

The cosmological parameters used in our fiducial model are:
\begin{center}
\begin{tabular}{lll}
$\Omega_m=0.3$, & $\Omega_\Lambda =0.7$, & $\Omega_bh^2 = 0.049$ \\
$\sigma_8=0.85$, & $h=0.7$, & $n=1$.
\end{tabular}  
\end{center}
Here, $n$ is the slope of the primordial power spectrum, which is 
filtered by the transfer function 
formulae from Hu and Eisenstein \cite{HuEisenstein}.  

We also need to specify the halo mass function $\bar n(m)$ for 
the chosen cosmology.
There are different prescriptions for obtaining $\bar n(m)$, the 
most well 
known being that of Press and Schechter \cite{PressSchechter}.  Two other 
and more 
accurate mass functions are widely used in the literature, namely that of
Sheth and Tormen \cite{ShethTormen}, and that of Jenkins \cite{Jenkins}.
We use 
the Sheth-Tormen halo mass function since it can be physically motivated 
using elliptical collapse.  The Sheth-Tormen halo mass function
is given by:
\begin{equation}
\bar n(m) = \frac{\Omega_m\rho_c}{m} f(\nu)  \frac{d\nu}{dm}
\label{eq:ST}\end{equation}
where 
\begin{equation}
f(\nu) =  A(1+(q\nu)^{-p})\big( \frac{q}{2\pi \nu} \big)^{\frac{1}{2}}
	  \exp(-\frac{q\nu}{2}).
\label{eq:STf}\end{equation}
The value $A=0.3222$
is obtained from numerical fits to $N$-body simulations, and recall from
\S II that $p=0.3$, $q=0.75$, and $\nu\equiv \delta_{sc}^2/\sigma^2(m)$.

\subsubsection{HOD}

We choose a halo occupation distribution in accord with our previous 
discussion, i.e. a host galaxy plus Poisson distributed satellite galaxies
for all halos. The average number of satellite galaxies is given by equation
(\ref{avgnumgal}) with  
\begin{center}
\begin{tabular}{cc}
$M_{1}=6.0\times 10^{12} M_\odot$ & \,\, $\alpha =1.0$. 
\end{tabular}
\end{center}

The choice of the HOD parameters requires a little discussion. 
Both of the HOD parameters  $\alpha$ and $M_{1}$ 
for large halo masses have been obtained 
empirically by doing halo model fits to the galaxy-galaxy correlation
function as measured by the SDSS (see Zehavi et al. \cite{Zehavietal}) and 
the 2dF survey (see Magliocchetti and Porciani \cite{MagliocchettiPorciani}).
Zehavi et al. find a slope of $\alpha\approx 0.89$, while Magliocchetti
and Porciani find $\alpha\approx 0.9$ for old galaxies, and 
$\alpha\approx 0.6$ for star forming galaxies 
(see \cite{MagliocchettiPorciani} for details).  Finally, Berlind et al. 
\cite{Andreas} find $\alpha\approx 0.9$ on the basis of numerical simulations.
In all of these fits, however, no distinction between a central galaxy
and satellite galaxies was made in any of these studies. Kravtsov et al. 
showed that in their N-body simulations this
distinction gives better fits in N-body simulations, while raising the slope
from 0.9 to 1.0 \cite{AndreyAndreas}.
We have therefore opted to use a slope of $1$ as the fiducial model
in accord with dark matter simulations.\footnote{We note here that 
baryon cooling may lower the value of $\alpha$. For instance,
in massive halos, the cooling time may approach 
the hubble time, which may reduce galaxy formation efficiency.
However, since we are not aware of either observational or galaxy
formation simulation constraints on the halo occupation distribution
where a distinction between host and satellite galaxies is made,
we have opted to keep the value of $\alpha$ obtained from 
dark matter simulations.}  Note that 
there is evidence that the slope varies with galaxy type  
(early vs. late, \cite{MagliocchettiPorciani}), so clearly
the correct value of the slope will depend on the exact sample of
galaxies we are looking at.

\begin{figure}[t]
\sfig{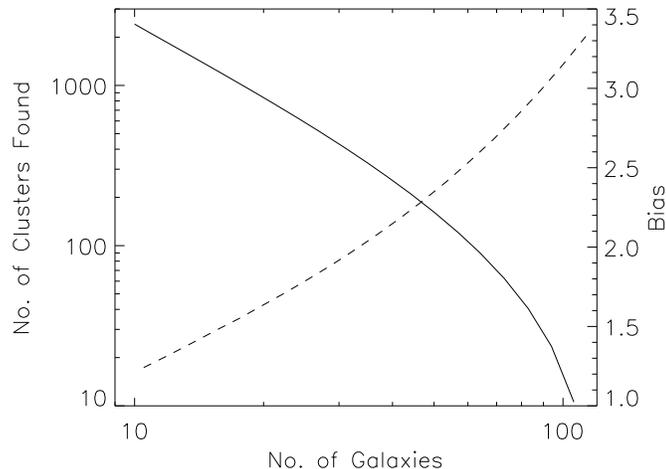}{0.55\columnwidth}
\caption{\footnotesize We show above the predicted richness (solid line) and
observed bias as a function of richness (dashed line)
for our survey given our fiducial model. The richness function is defined 
as the number of clusters found having more than the specified minimum 
number of galaxies.}
\label{observables}
\end{figure}

The mass parameter $M_1$ likewise depends on the particular galaxy sample
under consideration.  Most importantly, it depends critically on the
intrinsic luminosity cutoff of the sample.  For galaxies with intrinsic
luminosity $M_r<-21$, Zehavi et al. find the value
$M_1\approx 1.0\cdot 10^{14} M_\odot$.\footnote{The value $M_1$ they quote
is lower, but we have corrected it to take into account the assumption of 
a central galaxy.  Zehavi quotes that at a mass of 
$\approx 4.5 \cdot 10^{14} M_\odot$, one expects 5.4 galaxies, corresponding
to 4.4 satellite galaxies, or $M_1 \approx 1.0 \cdot 10^{14} M_\odot$ if
$\alpha=1$.}  The magnitude limit above
corresponds to a galaxy density $\approx 10^{-3} h^3$ gal/Mpc$^{3}$.
 Kravtsov et al. find a similar value for $M_1$ at said 
density.  For a density $n=2.79 \cdot 10^{-2} h^3$ Mpc$^{-3}$, corresponding
to galaxies brighter than $M_r \approx -18$ 
(Blanton et al. \cite{Blantonetal}), 
Kravtsov et al  obtain
$M_1 = 5\cdot 10^{12} M_\odot$ \cite{AndreyAndreas}.  
Using semi-analytical and SPH simulations, Berlind et al \cite{Andreas}
find $M_1 \approx 7 \cdot 10^{12} M_\odot$ for a comparable galaxy 
density.\footnote{Note what we are calling
$M_1$ here corresponds to $M_{\rm crit}$ in the Berlind et al. paper, i.e.
the amplitude of $\avgN$ at high masses.}
We assume here a value $M_1=6\cdot 10^{12} M_\odot$ for the fiducial model.

\subsection{Survey Assumptions}

We assume a $10^4$ deg$^2$ sky survey 
which is the target SDSS coverage \cite{Yorketal},
and a volume limit of $z\leq 0.1$. 
In the fiducial cosmology, this corresponds to a volume of about 
$6 \cdot10^{7}$ Mpc$^3$. At our volume limit of $z=0.1$,
and for the SDSS telescope, one expects
all galaxies with intrinsic magnitude $M\sim-18$ to be detected and have
their redshift 
measured\footnote{See http://www.sdss.org/documents/goals.html.}
justifying our choice for $M_1$ in the fiducial model.
Notice that while the volume $V$ has some error $\Delta V$ 
arising from the fact that redshifts have some intrinsic error,
we have set this term to zero assuming spectroscopic redshift of the 
galaxies is available.  This is reasonable given our very shallow survey,
though the error $\Delta V$ will become non-negligible in higher redshift 
surveys.


Finally, we assume all clusters in the
catalogue are binned logarithmically into 20 different bins.  We have checked
this binning is fine enough to accurately contail all the information in our
survey. The lowest and highest richness classes we consider are clusters with
10 and 120 galaxies respecitively, which corresponds to a mass range between
$M_{min}\sim 5\cdot 10^{13} M_\odot$ and $M_{max}\sim 7\cdot 10^{14} M_\odot$
The predicted cluster richness function is shown in figure \ref{observables}.

One final note: we are considering bias to be our observable.  To obtain the
bias, one needs to fit the correlation function as a whole as 
$\xi=b^2\xi_{LIN}$ (note $\xi_{LIN}$ varies with cosmology).  
The fit gives the bias measurement, while
the shape information is discarded.  We expect then that the confidence
regions we obtain here may be improved upon when the full information of
the correlation function is used.

\section{Results}

\subsection{Determination of the Confidence Regions}

Now that we have the correlation matrix of our observables, we estimate
the Fisher matrix for the HOD and cosmological parameters 
as (see e.g. \cite{Scott})
\begin{equation}
F_{ab}=\sum_{ij} (C^{-1})_{ij} \frac{\partial O_i}{\partial \lambda_a}
			       \frac{\partial O_j}{\partial \lambda_b}
\end{equation}
where $\lambda_a$ stands for the various parameters of interest and $O_i$
label our various observables.  As explained in \cite{Scott}, the matrix
$F$ serves as an approximation of the inverse correlation matrix for
the parameters $\lambda_a$, which we use to compute the confidence regions
reported here.  Our observables in this case are the number of clusters
found in each bin and their bias, and the parameters of interest are
those specifying the cosmology and the HOD. The one caveat is that, to
obtain the confidence regions, we will use as our parameters not the
model parameters themselves (i.e. $\alpha,M_1,\sigma_8,$...) but
their logarithms.  There are two motivations behind this:

\begin{itemize}
\item Using the natural logarithms enforces positivity in all parameters.
\item Traditionally, cluster surveys have been used to obtain
      constraints of the form $\sigma_8\Omega_m^\gamma=constant$.  These type 
      of constraints are obtained as eigenvectors of the fisher matrix when
      the natural logarithm of the parameters of interest are used in
      computing the fisher matrix, so using logarithms should make comparison
      to previous work more straightforward.
\end{itemize}

Unless stated otherwise, all of our confidence regions and intervals will
be marginalized over $h,n,$ and $\Omega_bh^2$ using gaussian priors

\begin{center}
\begin{tabular}{ccc}
$\sigma_h = 0.1$ & \hspace{0.2 in} $\sigma_n = 0.1$ \hspace{0.2 in} &
$\sigma_{\Omega_bh^2}=0.002$. 
\end{tabular}
\end{center}

\subsection{$M_1-\Omega_m$ Degeneracy}
\label{secm1om}

\begin{figure}[t]
\sfig{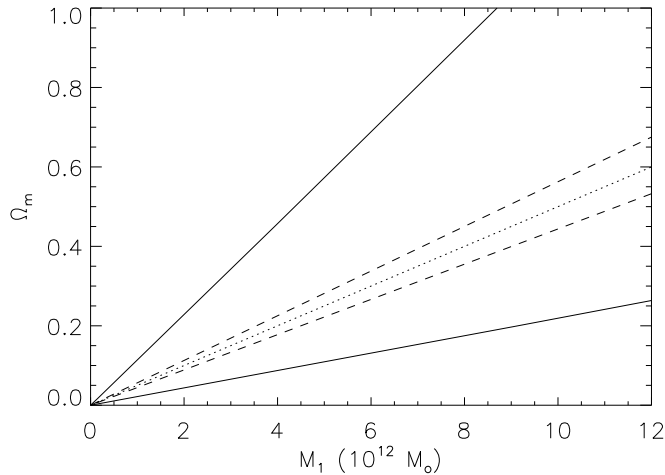}{0.55\columnwidth}
\caption{\footnotesize  The solid lines in this figure delimit the $95\%$ 
confidence region marginalized over all parameters.  Even moderately 
strong priors of
$\Delta \alpha=5\%$ and $\Delta \sigma_8=0.1$ do not improve the 
constraints much, though holding them fixed collapses the confidence
region to that enclosed by the dashed lines.  Finally,
the dotted line corresponds to the contour $M_1/\Omega_m=constant$,
the constant being set by the fiducial model.  The $95\%$ confidence
limits on $\Omega_m M_1^{-1}$ are 
$\Omega_m M_1^{-1} = 5.0^{+4.8}_{-2.5}\cdot 10^{-14} M_\odot^{-1}$ 
(marginalized over $\alpha,\sigma_8$) and 
$\Omega_m M_1^{-1} = 5.00^{+0.52}_{-0.47}\cdot 10^{-14} M_\odot^{-1}$ 
($\alpha,\sigma_8$ fixed).}
\label{om_m1_degeneracy}
\end{figure}

The first thing we notice upon computation of the Fisher matrix
is the existence of an extremely large degeneracy between $M_1$
and $\Omega_m$ of the form $\Omega_m/M_1=constant$.
This is shown graphically in figure \ref{om_m1_degeneracy}.  Here,
we plot the $95\%
$ confidence region in the $\Omega_m-M_1$ plane
when marginalizing over $\alpha$ and $\sigma_8$ (solid lines) and
when holding them constant (dashed lines).  The dotted line 
corresponds to the equation $\Omega_m/M_1= constant$, the constant 
being set by the fiducial model.  The $95\%
$ confidence constraint
marginalized over all other parameters is 
$\Omega_m/M_1 = (5.0^{+4.8}_{-2.5})\cdot 10^{-14} M_\odot^{-1}$, a rather
poor constraint.  The perpendicular direction 
$\Omega_m M_1=constant$ is not constrained at all.

The reason for this degeneracy can be traced to the 
behavior of the mass function with $\Omega_m$ and the fact that our
observable (number of galaxies in a cluster) scales as $m/M_1$.  
To see this mathematically, first note that for the halo mass function of 
Eq.~\ref{eq:ST},
\begin{equation}
dm \bar n(m,\Omega_m) \approx dx F(x)
\label{start}
\end{equation}
for $x=m/\Omega_m$ and $F$ being some function. 
Equation \ref{start} can be verified by referring back to equation \ref{eq:ST}.
We see there that $\bar n(m,\Omega_m)$ takes the form 

\begin{eqnarray}
dm \bar n & = & (\Omega_m dx)\frac{\rho_c}{x} \frac{d\nu}{dm}f(\nu)\\
	& = & dx \frac{\rho_c}{x}\frac{d\nu}{dx}f(\nu).
\end{eqnarray}

where $\nu=\delta_sc^2/\sigma(R(m))^2$.  Since $R(m)$ is given by the 
condition $4\pi R(m)^3 \Omega_m \rho_c = 3m$, it is clear that $R(m)$ 
depends only on $x$. Thus, if the power spectrum were independent of
$\Omega_m$, $\sigma(R(m))$, and hence $\bar n$, would depend only on $x$.
There is then a small dependence of $\sigma (R(m))$ on $\Omega_m$ alone
which comes through the power spectrum used in the convolution.  This
small dependence makes equation \ref{start} only approximate.
\\

Now, for a given mass $m$, the number of galaxies in the halo -- and therefore
the value of $\avgpsia$ -- depends via Eq.~\ref{avgnumgal} only on 
$(m/M_1)$.  That is, $\avgpsia(m)=g(m/M_1)$ for some function
$g$, so
\begin{eqnarray}
\bar\na & = & \int dm \bar n(m,\Omega_m) g(m/M_1) \nn \\
& \approx & \int dx F(x)g(\Omega_m x/M_1) \nn \\
& = & \phi(\Omega_m/M_1) 
\end{eqnarray}
where $\phi$ is a function.  We see then that $\bar\na$ depends
only the ratio $M_1/\Omega_m$, so there is indeed a full degeneracy 
between the two parameters.
Since the argument assumes only that the mass tracer scales as a function of 
$m/M_1$ where $M_1$ is a characteristic mass scale, we expect 
this degeneracy to be quite general.  It is also easy to check that a 
similar computation holds for bias.

\begin{figure}[!t]
\sfig{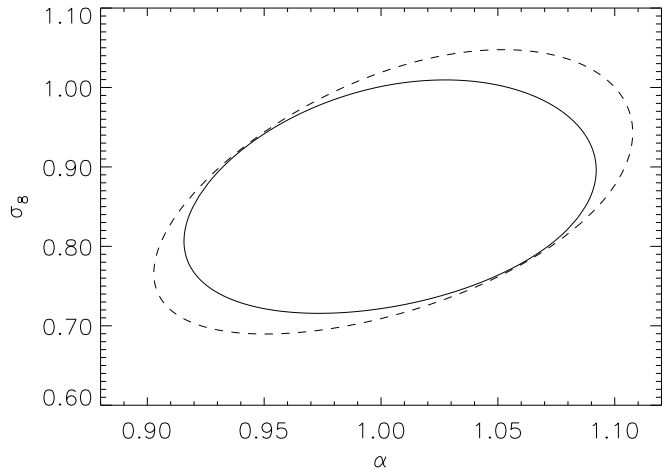}{0.55\columnwidth}
\caption{\footnotesize The $95\%$ confidence regions
marginalized over the Hubble rate, spectral index, and baryon density
using moderate gaussian priors.  The outer dashed ellipse is also
marginalized over $M_1$ and $\Omega_m$ without assuming any priors for
these quantities.
The inner solid ellipse is obtained assuming gaussian priors 
$\Omega_m = 0.3\pm 0.1$ and $\Delta M_1/M_1=30\%$.
The corresponding $95\%$
confidence intervals are  $\alpha=1.000^{+0.085}_{-0.079}$ 
and $\sigma_8=0.85^{+0.16}_{-0.13}$, and
$\alpha = 1.000^{+0.073}_{-0.068}$ and $\sigma_8=0.85^{+0.13}_{-0.11}$.}
\label{slope_s8}
\end{figure}

\subsection{Constraints on $\alpha$ and $\sigma_8$}

Despite the strong degeneracy between $M_1$ and $\Omega_m$, 
it is still possible to
constrain $\alpha$ and $\sigma_8$ without the use of any further
priors.  Shown in figure \ref{slope_s8} with the dashed line is the
$95\%
$ confidence region of the $\alpha-\sigma_8$ plane marginalized over 
$\Omega_m$ and $M_1$ assuming no priors for either of these two
quantities.  The constraints improve only slightly when invoking
priors on either $\Omega_m$ or $M_1$. 
Finally, the solid contour encloses the $95\%$ 
confidence region with reasonable priors 
$\Delta\Omega_m=0.1$ and $\Delta M_1/M_1=30\%$.

\begin{figure}[t]
\sfig{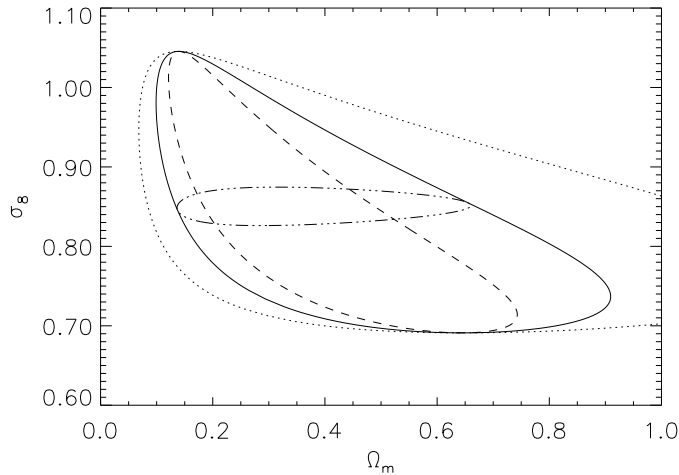}{0.55\columnwidth}
\caption{\footnotesize The $95\%$ confidence regions above are
marginalized over the hubble rate, spectral index, and baryon density
using moderate gaussian priors. In addition, we have assumed a gaussian prior
for the $M_1$ parameter of width $\Delta M_1=50\%$ (dotted line),
$\Delta M_1 =30\%$ (solid line), and $\Delta M_1=15\%$ (dashed
line).  The inner curve (dash-dot) is obtained by fixing $h$ and using the
$30\%$ prior on $M_1$.}
\label{s8_om}
\end{figure}

\subsection{The $\sigma_8-\Omega_m$ Plane}

We have seen it is impossible to obtain information about
$\Omega_m$ and $M_1$ simultaneously due to a degeneracy of the form
$\Omega_m /M_1=constant$.  For our fiducial model, we saw in particular
that $M_1$ and $\Omega_m$ satisfy 
$\Omega_m/M_1 = (5.0^{+4.8}_{-2.5})\cdot 10^{-14} M_\odot^{-1}$ 
($95\%
$ confidence).
The fact that 
$\Omega_m/M_1$ is so poorly constrained is reflected in
the constraints one may obtain on $\Omega_m$ when assuming priors
on $M_1$ and vice-versa.  
It seems then that local cluster abundances are
not well suited to constrain either $\Omega_m$ or $M_1$.

Up to date, most of the work on cluster abundances has been aimed
at providing confidence regions in the $\sigma_8-\Omega_m$ plane 
(see e.g. \cite{Pierpaolietal},\cite{Viana03},\cite{Allenetal},
\cite{Ikebeetal} and references therein).
It is helpful then
to study the type of constraints we can place in the $\Omega_m-\sigma_8$
plane, not only to touch base with other work in this area, but also
to see whether we can indeed expect to constrain cosmology with
local cluster surveys.

We show in figure \ref{s8_om} our constraints, where we plot the 
$95\%
$ confidence regions in the $\Omega_m-\sigma_8$ plane
for various gaussian priors on $M_1$ (marginalization over all other
parameters is also done just as before).  The dotted line is for 
a gaussian prior with $\Delta M_1=50\%
$, the solid line is
for $\Delta M_1 = 30\%
$, and the dashed line for $\Delta M_1=15\%
$.
The most strongly constrained combination of $\sigma_8$ and $\Omega_m$ using
a $30\%
$ prior on $M_1$ is 
$\sigma_8\Omega_m^{0.13}\approx const$.  Notice this constraint
is rather different from the one typically found in the literature 
from local cluster abundances, which looks more like 
$\sigma_8\Omega_m^\gamma\approx constant$ with 
$\gamma\sim 0.5$.
This is not surprising, however, given that there are major differences 
between our analysis and most previous treatments, 
including the use of clustering properties (bias), the 
probing of lower masses in the halo mass function, and a marginalization
over $h$, which is often fixed in cluster abundance 
analysis.
In light of this last point, we also plot in figure \ref{s8_om} the 
expected confidence region obtained when $h$ is fixed and a $30\%
$
gaussian prior on $M_1$ is used (dash-dot curve).  It should be evident that 
marginalization over $h$ is extremely important, and that constraints
derived by holding $h$ fixed are over-optimistic.

The main point that should be very clear from figure \ref{s8_om} 
is that, because of the
very strong $\Omega_m-M_1$ degeneracy, the constraints that we can
place in $\Omega_m$ are entirely determined by how well can the
amplitude of the mass-richness relation be calibrated. 
Further, we expect this to be a generic feature of all cluster abundance
studies (regardless of the mass tracer used) since we believe the
degeneracy stems from how the halo mass function scales with 
$\Omega_m$.

\subsection{Constraining Cosmology With Cluster Statistics}

We have derived above the various constraints on the $\sigma_8-\Omega_m$
plane that we expect to obtain from our model survey.  Importantly,
we saw that marginalization over $h$ is necessary to avoid
overly optimistic constraints.  In light of this sensitivity to $h$, it
seems worthwhile to determine what combination of $\sigma_8,\Omega_m,$ and 
$h$ is most strongly constrained by the data (assuming a 
prior calibration of $\alpha$ and $M_1$). 
We compute these combinations of parameters using a principal component
analysis of the estimated correlation matrix.  
We find then that, for gaussian priors
$\Delta M_1=30\%$ and $\Delta\alpha=10\%
$ we can express the cosmological
constraints ($95\%
$ confidence level) as

\begin{eqnarray}
\sigma_8 h^{-0.73} & = & 1.103^{+0.025}_{-0.025} \\
\bigl(\sigma_8\Omega_m^{0.57}\bigr)^{0.74} h
	& = & 0.374^{+0.133}_{-0.098} \label{cosm2} \\
\sigma_8^{-0.20}\Omega_m h^{-0.28} & = & 0.34^{+0.91}_{-0.25}
\end{eqnarray}

Not unexpectedly, the best constrained mode is mostly $\sigma_8$, but
with an important contribution from $h$ (if this were not the case, 
marginalization over $h$ would have had little effect before).  We can
see as well that $\Omega_m$ is the most poorly determined parameter, which
makes sense given the large assumed uncertainty in $M_1$ and the strong
$M_1-\Omega_m$ degeneracy.  The constraints
on $h$ that one can obtain with this method are at a moderately interesting 
level. If one introduces a prior $\Delta\Omega_m=0.1$,
the $68\%
$ confidence interval obtained from cluster 
statistics is $h=0.7\pm 0.09$, comparable to 
$h=0.72\pm 0.02 \pm 0.08$ from supernovae 
(Freedman et al. \cite{Freedmanetal}) or 
$h=0.72\pm 0.05$ from the CMB (assuming a $\Lambda$CDM cosmology, see e.g.
Spergel et al. \cite{WMAP}).

In view of the above comments, it is clear that aside from consistency checks,
the most important contribution from local cluster statistics for the
purposes of determining cosmological parameters is the 
accurate determination of $\sigma_8$.  In particular, using reasonable
priors for all other variables we expect it will be  possible to constrain 
$\sigma_8$ at the $\approx 10-15\%$ level with a $95\%
$ confidence level.\footnote{By reasonable priors we mean 
$\Delta \alpha/\alpha=10\%, \Delta M_1/M_1 = 30\%, \Delta \Omega_m=0.1,
\Delta h=0.1,$ and $\Delta n=0.1$.}

\begin{figure}[t]
\sfig{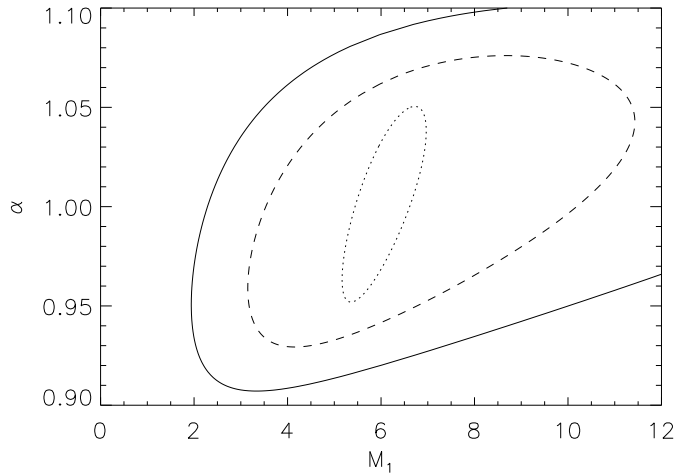}{0.55\columnwidth}
\caption{\footnotesize The $95\%$ confidence regions in the $\alpha-M_1$
plane.  Moderate gaussian priors 
$\Delta_m = \Delta n = \Delta h = 0.1$ and $\Delta \sigma_8 = 0.20$ were used
for the solid line, while the dashed line uses 
$\Delta \Omega_m = \Delta n = \Delta h =0.05$ and $\Delta \sigma_8=0.15$.
The inner, dotted ellipse fixes cosmology.  The $95\%$ confidence regions
for $\alpha$ when moderate priors are used is 
$\alpha=1.000^{+0.081}_{-0.075}$. 
If one keeps cosmology fixed, $M_1$ can also be constrained, and the
corresponding $95\%$ confidence regions are 
$M_1=6.00^{+0.76}_{-0.68}\cdot 10^{12} M_\odot$
and $\alpha=1.000^{+0.040}_{-0.039}$.}
\label{slope_m1}
\end{figure}

\subsection{Constraining the Halo Occupation Distribution}

Let us consider now the complementary problem of constraining the
halo occupation distribution by either fixing cosmology or marginalizing
over it.  These constraints should be of great interest in that 
they may help guide theoretical efforts in galaxy formation models.
Our results are shown in figure \ref{slope_m1}, where we plot the 
$95\%
$ confidence regions on the $\alpha-M_1$ plane when we hold
cosmology fixed (dotted line), marginalizing with strong
gaussian priors $\Delta \Omega_m,\Delta n,\Delta h =0.05$ and 
$\Delta \sigma_8=0.15$ (dashed line), and marginalizing over
moderate priors $\Delta \Omega_m,\Delta n,\Delta h = 0.1$ and 
$\Delta \sigma_8 = 0.20$.  As we expected, $M_1$ is poorly constrained
and its confidence interval depends on the prior used in $\Omega_m$.
The slope $\alpha$ on the other hand, can be well constrained.  Our
moderate priors lead to a $95\%
$ confidence interval
$\alpha=1.000^{+0.081}_{-0.075}$.  If one insists in keeping cosmology
fixed, however, both $\alpha$ and $M_1$ may be constrained to a good
accuracy.  In particular, for fixed cosmology the $95\%
$ confidence
regions become $\alpha=1.000^{+0.040}_{-0.039}$ and 
$M_1=6.00^{+0.76}_{-0.68}\cdot 10^{12} M_\odot$.

\begin{table}[!t]
\begin{center}
\begin{tabular}{|c|c|c|c|}
\hline
Assumptions	& $\Delta \alpha$ & $\Delta \sigma_8$ & $\Delta h$ \\ \hline
$\Delta\Omega_m,\Delta n = 0.1$	
                & $\pm 0.038$	  & $\pm 0.079$	      & $\pm 0.089$ \\ \hline
$\Delta\Omega_m,\Delta n = 0.05$
		& $\pm 0.032$	  & $\pm0.048$	      & $\pm 0.070$ \\ \hline
$\Delta\Omega_m,\Delta n,\Delta h = 0.1$
		& $\pm 0.036$  & $\pm0.080$	      & - \\ \hline
$\Delta\Omega_m,\Delta n,\Delta h = 0.05$
		& $\pm 0.028$ & $\pm 0.037$	      & - \\\hline
\end{tabular}
\end{center}
\caption{\footnotesize 
$1-\sigma$ predictions for $\alpha,\sigma_8$, and $h$ from 
cluster samples obtained with SDSS type surveys.  All error bars assume
gaussian priors $\Delta \Omega_bh^2=0.002$, and 
$\Delta M_1=30\%
$ as well as the assumptions listed on the table.}
\label{predictionstable}
\end{table}

\subsection{What Do Local Cluster Abundances and Bias Tell us?}
\label{hod_plus_cosm}

Given our above results, it seems fair to ponder as to 
what the best use of local cluster samples is.  We have seen
that bias allows us to constrain $\alpha$ and $\sigma_8$ simultaneously,
though large degeneracies with $h$ exist.  We have also seen that 
cluster abundances and bias can only constrain the combination 
$M_1/\Omega_m$, and only poorly at that.  It seems then that rather
than attempting to constrain cosmology alone or the mass tracer relations
alone, the data is best used to constrain the three parameters
$\alpha,\sigma_8,$ and $h$.

We quote in table \ref{predictionstable} how well 
can we constrain the various 
parameters ($\alpha,\sigma_8,$ and $h$) under various assumptions.  This
is meant to illustrate the power of cluster samples obtained from 
SDSS type surveys.  We assume in all cases gaussian priors
$\Delta \Omega_b h^2=0.002$, and $\Delta M_1=30\%
$.  
The values listed under the assumptions column are the values used
as gaussian priors for the appropriate variable.  All confidence
intervals are $68\%
$ and marginalized over all other parameters.

Is this the best we can do? Yes and no. On the one hand, we can perform a
singular value decomposition of the Fisher matrix assuming no priors
on any of the parameters, thus  determining which and how many directions 
are strongly constrained given our assumptions.
We find that there are 
indeed three directions which may be strongly constrained, and these
are most closely aligned to the $\alpha,\sigma_8$ and $h$ subspace. 
Thus, we cannot hope to place more than three strong constraints,
and if we want to choose three of our parameters, our best choice is
the triplet $(\alpha,\sigma_8,h)$ considered above.  Nevertheless,
the three directions which are most strongly constrained all have 
some contribution from the parameters $M_1/\Omega_m$ and $n$.  In
principle, then, we could do ``better" if we opt  for constraining
combinations of the five parameters 
$(\alpha,\sigma_8,h,M_1/\Omega_m,n)$, though only three such combinations
may be constrained.

\section{Cluster Abundances}

\subsection{The Role of Bias as a Complement to Cluster Abundances}

We wish to consider now the type of constraints we can place if we 
have cluster abundance information alone.  Before we do this, it 
is important to address whether the inclusion of clustering properties 
(in the form of cluster bias) contains information not included in
the cluster richness function.  If not, repeating our
analysis disregarding bias information would not alter any of our results.
We demonstrate that bias does indeed carry some additional information 
by considering how well can $\alpha$ be determined from our data.

\begin{figure}[t!]
\sfig{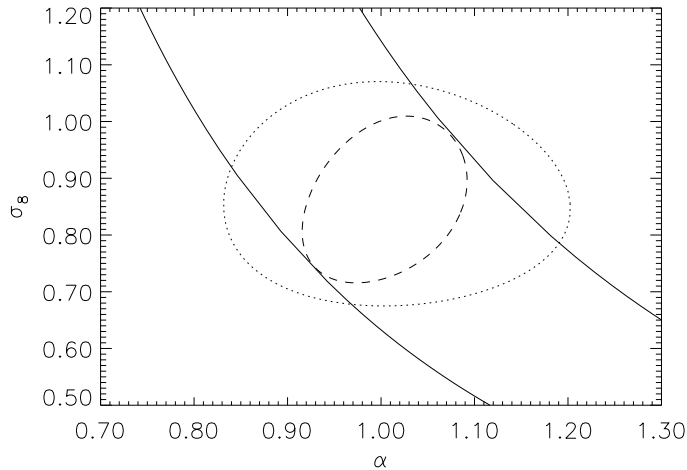}{0.55\columnwidth}
\caption{\footnotesize The $95\%$ confidence regions in
the $\alpha-\sigma_8$ plane, marginalized over $h,n,\Omega_b,M_1,$
and $\Omega_m$ with the usual priors on $h,n,$ and $\Omega_b$, and
gaussian priors $\Delta M_1=30\%$ and $\Delta\Omega_m=0.1$ for the
last two parameters.  The solid line is obtained when using cluster
abundances but no bias information, while the dotted line corresponds
to the converse case. We see that bias breaks an $\alpha-\sigma_8$ 
degeneracy which exits when using only cluster abundance information. 
Shown for comparison, the dashed ellipse above 
is the same as the dashed ellipse in figure \ref{slope_s8}, 
i.e. the confidence region obtained with both
cluster abundance and bias information with the aforementioned priors
on $M_1$ and $\sigma_8$.}
\label{slope_s8_nobias}
\end{figure} 

Figure \ref{slope_s8_nobias} shows the $95\%
$ confidence
regions in the $\alpha-\sigma_8$ plane, marginalized over all other 
parameters, and 
where in addition to the usual priors on $h$,$n$, and $\Omega_b$ we
used priors $\Delta M_1=30\%
$ and $\Delta\Omega_m=0.1$.  The solid
lines are obtained by using cluster abundances but ignoring bias, while
the dotted lines are obtained by ignoring cluster abundances but 
including bias.  Also shown as a dashed ellipse is the curve we obtain when
both bias and abundances are used.  Note this last ellipse is the same 
as the dashed ellipses in figure \ref{slope_s8}.
The important result then is that cluster abundances are degenerate in 
$\alpha-\sigma_8$ ($\alpha\sigma_8^{0.47} \approx constant$)
while bias breaks this degeneracy. This  demonstrates explicitly that
information contained in the bias complements that of cluster
abundances by themselves.

An important consequence of this argument is that, when bias
information is not included, one needs to either fix $\alpha$ or 
assume some prior on it.  We checked this explicitly by noting that the
$\sigma_8-\Omega_m$ confidence regions obtained from cluster abundance 
alone blow up if no prior is placed on $\alpha$.

\subsection{Can Local Cluster Abundance Alone Constrain
            $\sigma_8$ and $\Omega_m$?}

\begin{figure}[t]
\sfig{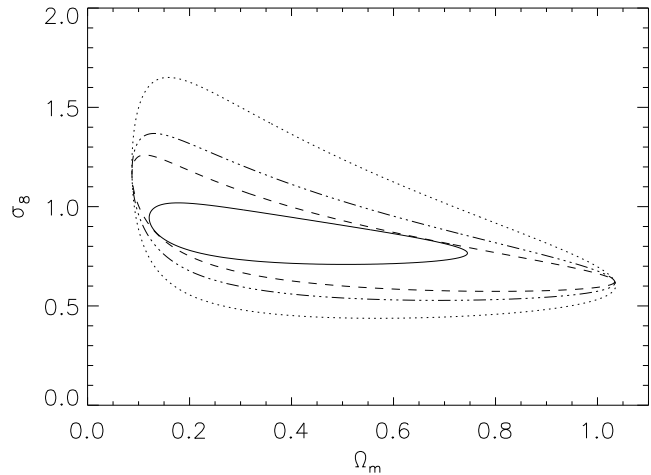}{0.55\columnwidth}
\caption{\footnotesize The $95\%$ confidence regions above are obtained
using gaussian priors on $\Omega_b,n,$ and $h$, and disregarding all 
bias information. i.e. these are the expected confidence regions from
optical cluster abundance studies.  Note in particular the scale
on the $\sigma_8$ axis.
The various curves correspond
to different priors on the richness-mass relation parameters.  These 
are $\Delta\alpha/\alpha =10\%$ (dotted), 
$\Delta\alpha/\alpha =5\%$ (dash-dot), and
$\Delta\alpha=0$ (dashed).  
All contours are obtained using a $30\%$ gaussian prior on $M_1$.  For
comparison, we also
show the contour obtained when we add bias information and keep $\alpha$
fixed, shown here in the solid line. Compare also to figure \ref{s8_om}
(though note the change in scale for the $\sigma_8$ axis).}
\label{s8_om1}
\end{figure}

In view of the degeneracy between $\alpha$ and 
$\sigma_8$ in cluster abundance considerations, 
we wish to consider to what extent  the 
$\sigma_8-\Omega_m$ plane can be constrained using cluster abundances alone.
We have already seen that due to the $M_1-\Omega_m$ degeneracy, the
constraint on $\Omega_m$ is determined in its entirety by how well
can the amplitude of the richness-mass relation be calibrated.  
Likewise, we expect that the 
constraints on $\sigma_8$ derived from cluster abundance studies to be
entirely determined by the calibration constraints on $\alpha$.
This is shown explicitly in figure \ref{s8_om1}, where we plot the 
$95\%
$ confidence regions in the $\sigma_8-\Omega_m$ plane for three
different gaussian priors on $\alpha$: $\Delta\alpha=10\%
$
(dotted line), $\Delta\alpha=5\%
$ (dash-dot), and
$\alpha=constant$ (dashed line).  A gaussian prior $\Delta M_1=30\%
$
is used, and we marginalized over $\Omega_bh^2,n,$ and $h$ with the usual
priors.   When computing these confidence regions, 
we disregarded all bias information.  Also shown for reference in the solid
line is the contour obtained when both bias and cluster abundances are used
and assuming a fixed $\alpha$. We note that when including bias information, 
the confidence $\sigma_8-\Omega_m$ region is not considerably worsened by
letting $\alpha$ float (not shown).

It should be clear from figure \ref{s8_om1} that, as we expected,
the $\sigma_8$ confidence interval is
entirely determined by the prior on $\alpha$, just as the $\Omega_m$
interval is determined by the $M_1$ prior.  Further, we note
that the constraints on $\sigma_8$ are rather weak for realistic
priors on $\alpha$.  In particular, even when all parameters except
$\alpha$ and $\sigma_8$  are fixed, the $95\%
$ confidence interval assuming
gaussian priors $\Delta \alpha=10\%
$ and $\Delta \alpha=5\%
$ are
$\sigma_8=0.85^{+0.38}_{-0.26}$ and $\sigma_8=0.85^{+0.19}_{-0.15}$
respectively.

There are thus two main results
from this exercise. Firstly, cosmology can be constrained by optical
cluster abundances only to the extent that a careful calibration of
the richness-mass relation can be achieved.  Secondly, we find that  
using realistic assumptions for the accuracy of the calibration parameters
neither $\sigma_8$ nor $\Omega_m$ may be accurately constrained by cluster
abundances alone. As a corollary, it is evident that
any constraints placed on $\Omega_m$ and $\sigma_8$ using cluster abundances
which are not properly marginalized over $M_1$ and $\alpha$ will lead to 
overly optimistic constraints (see figure \ref{s8_om2}).

\subsection{What do Local Cluster Abundances Alone Tell Us?}
\label{what_abundances_tell_us}

As in section \ref{hod_plus_cosm}, we can now ask ourselves how we can
best use cluster abundance information.  We attack this question by finding
the eigenvalues and eigenvectors of the estimated parameter correlation 
matrix.  Upon doing so, we see that there are only two directions which
are strongly constrained.  These two eigenvectors are most closely
aligned with the $(\alpha\sigma_8^{0.5}-h)$ plane, but have contributions 
from 
$\Omega_m/M_1$ and $n$, as well as a very weak contribution from 
$\sigma_8$.  
For reference, we write the eigenvectors below\footnote{In 
these expressions we have disregarded the very weak dependence on $\sigma_8$
alone.  This dependence shows up as an extra factor of $\sigma_8^{0.03}$ in
both eigenvectors}

\begin{eqnarray}
\alpha \Bigl( \sigma_8 \bigl(\Omega_m/M_1\bigr)^{0.73}\Bigr)^{0.5}h^{0.51}
        + 0.15\delta n  \label{egv1} \\
\bigl(\alpha\sigma_8^{0.50}\bigr)^{0.66} \bigl(\Omega_m/M_1\bigr)^{-0.23}h^{-1}
	- 0.39\delta n \label{egv2}
\end{eqnarray}

where $n=1+\delta n$.
Though these may not seem terribly illuminating, these expressions
contain much information.  Consider the first eigenvector: if we take
all parameters except $\sigma_8$ and $\Omega_m$ to be fixed, this
eigenvector reduces to $\sigma_8\Omega_m^{0.73}=constant$, a cluster 
normalization condition.  Thus, equation \ref{egv1} may be thought of as 
a generalized cluster abundance normalization condition (see Appendix
\ref{appA} for more discussion).

The second eigenvector above, equation \ref{egv2}, does not have a 
simple interpretation (though see Appendix \ref{appA}).  Regardless,
there are still 
elements which are of interest. Importantly, only the combinations
$\Omega_m/M_1$ and $\alpha\sigma_8^{0.5}$ appear, which are the degeneracies
we have already found. This confirms that this degeneracies are indeed
intrinsic to cluster abundance studies and cannot be avoided.  

As a final note and in answer to the question posed by the title of this
section, we state here that cluster abundances are most well suited to
constrain the combination $\alpha\sigma_8^{0.50}$ and $h$.  The constraints
are somewhat sensitive to $n$ and $\Omega_m/M_1$, but only moderately so. 
We show in table \ref{abundancetable} the $68\%
$ confidence intervals
for $\alpha\sigma_8^{0.50}$ under various assumptions.

\begin{table}[t]
\begin{center}
\begin{tabular}{|c|c|c|}
\hline
Assumptions			& $\Delta \alpha\sigma_8^{0.5}$  \\ \hline
				& 0.090 \\ \hline
$\Delta\alpha=10\%
$		& 0.074 \\ \hline
$\Delta\alpha=10\%
,\Delta\sigma_8=0.2$ & 0.064 \\ \hline
$\Delta\alpha=10\%
,\Delta\sigma_8=0.2,\Delta h =0.1$ & 0.050 \\ \hline
*						     & 0.053 \\ \hline
* + $\Delta h=0.05$				     & 0.035 \\ \hline
\end{tabular}
\end{center}
\caption{\footnotesize $1-\sigma$ predictions for $\alpha\sigma_8^{0.5}$ from 
cluster samples obtained with SDSS type surveys.  All error bars assume
gaussian priors $\Delta \Omega_bh^2=0.002$, $\Delta M_1=30\%, \Delta n =0.1,$ 
and $\Delta \Omega_m=0.1$, as well as the assumptions listed on the table.
The next to last row, marked with a * in assumptions, 
assumes $\Delta \Omega_m=0.05$
and $\Delta n=0.05$, as well as priors $\Delta\alpha=10\%$ and 
$\Delta\sigma_8=0.2$.  In all cases where no prior on $h$ is assumed,
the $1-\sigma$ interval for $h$ is $h=0.7\pm0.1$, which reduces to
$h=0.70\pm0.08$ for case *. }
\label{abundancetable}
\end{table}

\subsection{Origin of the Degeneracies}

Perhaps the most surprising result so far is that
local cluster abundances alone are not capable of constraining 
either $\Omega_m$ or $\sigma_8$ very precisely when using reasonable
priors for the richness-mass relation.  This result, however,
was obtained for optical cluster surveys, which amounts operationally
to a choice of mass scale $M_1$ and power law index $\alpha$ in the 
scaling of the mass tracer with halo mass.  It is possible then that our 
conclusions do not hold in the case of X-ray surveys or other mass tracers,
where one would have different values for $M_1$ and $\alpha$.
Here, we
wish to investigate where the degeneracies stem from to determine
whether we expect them to be generic or particular to 
the fiducial model we have assumed.

\subsubsection{$M_1-\Omega_m$ Degeneracy}

Let us begin by analyzing the $M_1-\Omega_m$ degeneracy first.
We concluded in section \ref{secm1om} that as long as the mass tracer
scales with mass as some function of $m/M_1$ for some characteristic
mass scale $M_1$, then the degeneracy between
$M_1$ and $\Omega_m$ will always exist.  The only possible way out of this
statement is that our starting assumption, namely equation 
\ref{start}, is violated.  We discuss then what conditions equation
\ref{start} imposes on the mass function.

Let us assume that equation 
\ref{start} holds and consider the product $dx f(x)$ for for two 
values $(m,\Omega_m)$ and $(m',\Omega_m')$.  Equation \ref{start}
implies
\begin{equation}
dm \bar n(m,\Omega_m) = dm' \bar n(m',\Omega_m')
\end{equation}
provided $m'/\Omega_m'=m/\Omega_m$.  Defining 
$\lambda\equiv\Omega_m'/\Omega_m$, we obtain $m'=\lambda m$,
which upon replacing on the right hand side above yields
\begin{equation}
dm \bar n(m,\Omega_m) = dm \lambda \bar n(\lambda m,\lambda \Omega_m).
\end{equation}

In other words, equation \ref{start} holds if and only if the halo
mass function satisfies the scaling relation\footnote{The converse
is proved simply by reversing our argument.}
\begin{equation}
\bar n(m,\Omega_m) = \lambda \bar n(\lambda m,\lambda \Omega_m).
\end{equation}
Letting $\Omega_m=1$, this relation simplifies to
\begin{equation}
\bar n(m)|_{\Omega_m=1} = \Omega_m \bar n(\Omega_m m, \Omega_m).
\end{equation}
which was indeed found by 
Zheng et al. \cite{Zhengetal} using extensive numerical simulations 
In fact, they found the correlation function scales in a similar way,
supporting the fact that the $M_1-\Omega_m$ degeneracy we found
is not broken when bias is included as an additional observable.

We are forced to conclude then that there is an intrinsic limit to how well 
can we constrain $\Omega_m$ from cluster abundances which is set by
the uncertainty in the characteristic mass scale $M_1$.  Mass measurements at
present are only accurate to about $20\%$ (optimistically) to $50\%
$ 
(pessimistically) , which essentially sets
the maximum accuracy one could achieve in $\Omega_m$ using local cluster 
surveys.  Note, however, that we found that the ratio $M_1/\Omega_m$
itself was rather poorly constrained, so it seems unlikely that
local cluster abundances can provide strong constraints on 
$\Omega_m$.\footnote{Note we are not making any such claim for
cluster abundance studies extending over a large redshift range,
since then it is possible to provide constraints using the observed
growth of structure.}

\subsubsection{$\alpha-\sigma_8$ Degeneracy}

Let us now turn to the $\alpha-\sigma_8$ degeneracy.   We perform 
an analysis similar to the one above to determine if the degeneracy stems
from a scaling property of the halo mass function.

Consider then a variation of 
$(\alpha,\sigma_8) \rightarrow (\alpha',\sigma_8')$
which leaves cluster abundances fixed. The relation between the
parameters  is easily obtained from a singular value decomposition 
of the Fisher matrix holding all parameters fixed except for $\alpha$ and 
$\sigma_8$.  We obtain that cluster abundances are approximately 
degenerate when $\alpha\sigma_8^{0.5}=\alpha'\sigma_8'^{0.5}$.

Now, since our mass tracer scales with mass as a power law, 
the binning functions may be expressed in terms of a function 
$g(\alpha\ln(m/M_1))$.  We have then 
\begin{equation}
\Na(\alpha',\sigma_8') \propto  
\int dm' \bar n(m',\sigma_8') g(\alpha'\ln (m'/M_1) ). \nn
\end{equation}

We perform now a change of variables by defining $m$ via 
$\alpha\ln(m/M_1) = \alpha'\ln(m'/M_1)$.  Defining 
$\lambda=\alpha/\alpha'$ and replacing above we get

\begin{equation}
\Na(\alpha',\sigma_8') \propto
\int dm \lambda (m/M_1)^{\lambda-1}\bar n(m',\sigma_8') 
     g(\alpha\ln(m/M_1)) 
\label{comp1} 
\end{equation}

which is to be compared with

\begin{equation}
\Na(\alpha,\sigma_8) \propto
\int dm \bar n(m,\sigma_8) g(\alpha\ln (m/M_1)).
\label{comp2}
\end{equation}

If we demand that $\sigma_8'=\lambda^2\sigma_8$, we have then that
$\alpha\sigma_8^{0.5} = \alpha'\sigma_8'^{0.5}$, and thus by construction
$\Na(\alpha,\sigma_8) \approx \Na(\alpha',\sigma_8')$.  Since the function
$g$ is the same in both equations \ref{comp1} and \ref{comp2}, this 
suggests that the kernel of the integrals are nearly degenerate.
We thus make the ansatz

\begin{equation}
\bar n(m,\sigma_8) \approx 
     \lambda \Bigl(\frac{m}{M_1}\Bigr)^{\lambda-1} 
     \bar n(M_1(m/M_1)^\lambda,\lambda^2\sigma_8).
\label{mfscaling}
\end{equation}

\begin{figure}[t]
\sfig{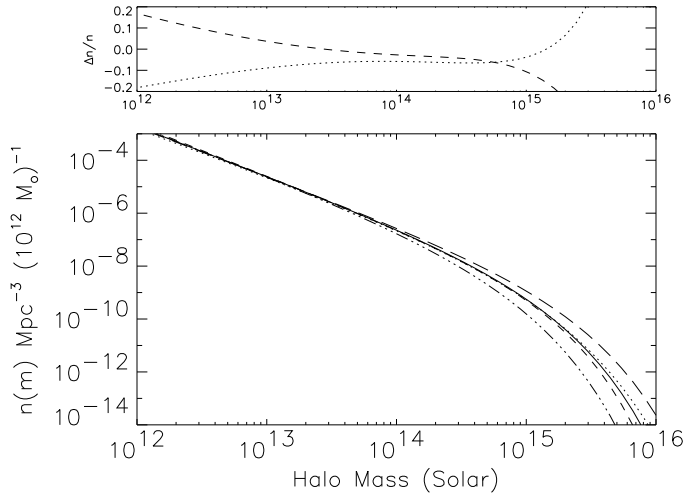}{0.55\columnwidth}
\caption{\footnotesize Shown above in the solid line is the Sheth Tormen
halo mass function for our fiducial model ($\sigma_8=0.85$).  
Also shown are the scaled
mass functions (i.e. right hand side of equation \ref{mfscaling}) for 
$\sigma_8'=1.0$ (dashed line) and $\sigma_8'=0.7$ (dotted line). For 
comparison, we also show the unscaled halo mass functions for $\sigma_8=1.0$
and $\sigma_8=0.7$ with the long-dash and dash-triple dot lines respectively.
The scaled mass functions are seen
to agree with each other to within $\approx 10\%$ or better 
in the mass range $10^{12}-10^{15} M_\odot$.  
The difference between the scaled mass function
and our fiducial model mass function, divided by the halo mass function, 
is shown in the upper panel.  Conventions are the same: $\sigma_8'=1.0$
corresponds to the dashed line while $\sigma_8'=0.7$ is shown with the
dotted line.}
\label{halomf1}
\end{figure}

We check our ansatz using the Sheth-Tormen halo mass function.  In particular,
figure \ref{halomf1} shows the Sheth-Tormen halo mass function, computed
for our fiducial model, as well as the scaled mass functions computed 
according to equation \ref{mfscaling}.  The smaller plot above shows the 
fractional difference between the right and left hand sides of equation
\ref{mfscaling}.  The right hand
side of equation \ref{mfscaling} was computed for 
$\sigma_8'=1.0$ (dashed line) and 
$\sigma_8'=0.7$ (dotted line).  We find that our scaling relations 
are accurate to within $\approx 10\%
$ for $\sigma_8$ in the range $[0.7,1.0]$ and over the mass range
$10^{13} M_\odot - 10^{15} M_\odot$.  Even though the accuracy of the 
scaling relations decreases as we move away from this range, it is still 
better than $20\%
$ from $10^{12} M_\odot$ up to 
$\approx 2\cdot 10^{15} M_\odot$ for $\sigma_8$ in the range
$0.7<\sigma_8<1.0$.

We conclude therefore that, in the mass ranges probed by our model, there
appears to exist a scaling relation given by equation \ref{mfscaling}.
Further, 
running the argument we used to derive our ansatz in reverse proves 
that the degeneracy we observe does indeed stem from the scaling relation 
\ref{mfscaling}.  We thus expect our results to hold for all 
mass tracers that scale with mass via a power law and which probe the
mass range $10^{13}-10^{15} M_\odot$.  In particular,
we expect a degeneracy between $\alpha$ and $\sigma_8$
to exist for all cases (though recall bias breaks this degeneracy) as
well as an $M_1-\Omega_m$ degeneracy.  This further implies that 
 any determination of 
cosmological parameters done using local cluster abundances alone
needs to be marginalized over uncertainties in how the mass tracer
scales with mass, regardless of the choice of mass tracer.  
Note marginalization
over the hubble rate is also necessary due to large mixing between the
cosmological parameters for fixed cluster densities.

To close, we note the fact that the scaling relation \ref{mfscaling}
depends on $M_1$.  This suggests that a more accurate
degeneracy can be achieved by allowing $M_1/\Omega_m$ to vary using for 
example equations \ref{egv1} and \ref{egv2}.
Nevertheless, the relation we obtained works well, and is only
weakly dependent on the value of $M_1$.   
Indeed, we can change $M_1$ by up to a factor
of two up or down and still get a reasonable agreement in the scaling
relation \ref{mfscaling}.   It is likely therefore that the role of
$M_1$ is essentially to set the mass scale over which the halo mass
function will satisfy scalings of the form \ref{mfscaling}.

\section{Conclusions}

We have derived expressions for cluster abundances and bias using the
halo model formalism. In particular, we have shown how starting from
the halo mass function and halo bias, we can obtain expressions for
cluster statistics in terms of any mass tracer (e.g. 
X-ray temperature/luminosity, number of galaxies, etc.) 
in such a way that
various experimental effects can be included in the formalism.  Specifically,
we include intrinsic dispersion of the richness-mass relation, bias and/or
scatter due to experimental measurements, detection rates, 
and false detections.  We also identified various sources of errors and
derived the uncertainties one expects due to intrinsic scatter in the 
richness-mass relation.  We believe that these derivations are important
in that they allow us to compare theory directly to observations,
without having to manipulate the observations in attempts to 
retrieve, for instance, the halo mass function.  Further, they allow us
to consider the possibility of using cluster statistics to constrain
not just cosmology, but also how the mass tracer scales with halo
mass.

Having derived our formalism, we applied it to the case of large local 
optical cluster surveys.  In this context,
we have shown that optical cluster survey determinations of cluster
abundances and bias can provide strong
constraints on the amplitude of the power spectrum at cluster scales 
($\sigma_8$),
the power law index on the scaling relation of the number of galaxies
in a halo of mass $m$, and perhaps even the hubble parameter $h$ 
(see table \ref{predictionstable}). We argued as well that cluster abundances
and bias are not well suited for constraining $\Omega_m$ or $M_1$, the
amplitude of the mass tracer scaling relation with mass.

We have shown as well that one needs to be very careful when analyzing
cluster abundances and bias data in order
to avoid overly optimistic constraints.  In particular, we have shown that
realistic constraints on $\sigma_8$ need to be marginalized over $h$ and,
when bias information is unavailable, 
marginalization over priors on $\alpha$ is also necessary.  
Though marginalization over
$M_1$ has a much smaller effect on $\sigma_8$, it becomes of paramount
importance if one wishes to constrain $\Omega_m$.  In fact, we found that
the uncertainties in $\sigma_8$ and $\Omega_m$ obtained from cluster
abundance studies alone are entirely determined by the priors used
on $\alpha$ and $M_1$.

We have attempted to explain why is it that the uncertainties in 
$\sigma_8$ and $\Omega_m$ are driven by the priors on $\alpha$
and $\Omega_m$ when only cluster abundance information is used.
In particular, we have argued that this effect is driven by
two degeneracies, one involving $M_1$ and $\Omega_m$, the other
involving $\alpha$ and $\sigma_8$.  We have shown here that these
degeneracies arise from scaling laws satisfied by the halo mass
functions.  The scaling law leading to the 
$M_1-\Omega_m$ degeneracy was found empirically by Zheng et al.
\cite{Zhengetal}, but it is reassuring to see it re-emerge here
in our Fisher analysis. The scaling law \ref{mfscaling} leading to the
$\alpha-\sigma_8$ degeneracy is, to the best of our knowledge,
a new result.

Finally, we argued that because the degeneracies above stem
from intrinsic scaling relations of the halo mass function, our 
conclusions are valid for any cluster abundance study in which
the mass tracer scales with mass as a power law to a good approximation.
In particular, for any such studies, any constraints that one wishes
to place on cosmology need to be properly marginalized over the 
hubble rate $h$ as well as the amplitude and power law index of the
mass tracer scaling relation.

In summary, then, we have found that cluster abundances and cluster bias 
are powerful tools that can greatly constrain both the amplitude of the
power spectrum at cluster scales, and how the number of galaxies in a
halo scales with mass.  However, neither $\Omega_m$ nor the 
characteristic mass scale for the formation of galaxies can be accurately
constrained.  In either case, it is always important to marginalize over
$h$ in order to avoid obtaining unrealistically tight constraints.

\section{Acknowledgments}

We would like to thank Andrey Kravtsov, Andreas Berlind, Risa Wechsler, 
and James Annis for helpful comments and discussion.  We would also like
to thank our referee for useful comments which greately improved our
presentation. 
This work was carried out at the University of Chicago, Center for 
Cosmological Physics and was suported in part by NSF PHY-0114422 and
NSF Grant PHY-0079251.

\newcommand\AAA[3]{{A\& A} {\bf #1}, #2 (#3)}
\newcommand\PhysRep[3]{{Physics Reports} {\bf #1}, #2 (#3)}
\newcommand\ApJ[3]{ {ApJ} {\bf #1}, #2 (#3) }
\newcommand\PhysRevD[3]{ {Phys. Rev. D} {\bf #1}, #2 (#3) }
\newcommand\PhysRevLet[3]{ {Physics Review Letters} {\bf #1}, #2 (#3) }
\newcommand\MNRAS[3]{{MNRAS} {\bf #1}, #2 (#3)}
\newcommand\PhysLet[3]{{Physics Letters} {\bf B#1}, #2 (#3)}
\newcommand\AJ[3]{ {AJ} {\bf #1}, #2 (#3) }
\newcommand\aph{astro-ph/}
\newcommand\AREVAA[3]{{Ann. Rev. A.\& A.} {\bf #1}, #2 (#3)}

\appendix

\section{Where Did the Usual $\sigma_8-\Omega_m$ Degeneracy Go?}
\label{appA}

Cosmological constraints from cluster abundance studies are usually 
expressed in the form of the so called cluster abundance normalization 
condition,
usually expressed as $\sigma_8 \Omega_m^\gamma \approx 0.5$ where 
$\gamma\approx 0.5$.  In other words, cluster abundances are usually used
to constrain the combination $\sigma_8 \Omega_m^\gamma$.  This combination
of parameters, however, did not arise from our analysis. Where did the 
usual $\sigma_8-\Omega_m$ degeneracy go?

\subsection{The $\sigma_8-\Omega_m$ Degeneracy}

Let us consider first equation \ref{egv1}.  We stated in section
\ref{what_abundances_tell_us} that \ref{egv1} could be though of
a generalized cluster abundance normalization condition.
While the exponent of 
$\Omega_m$ is a little steeper than usual, we show below that this
arises simply because we are probing rather low mass scales.

\begin{figure}[t]
\sfig{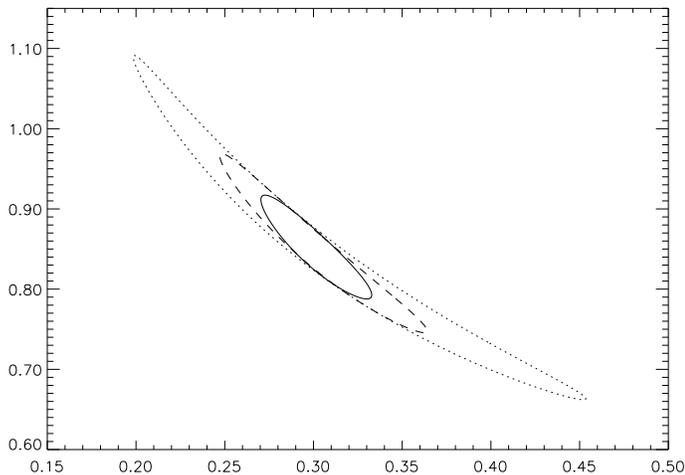}{0.55\columnwidth}
\caption{\footnotesize The figure above shows two things: first, that when
we restrict our analysis to $\sigma_8$ and $\Omega_m$ holding all other
parameters fixed, and considering only massive clusters ($N_{gal}\geq 35$)
we recover the usual $\sigma_8-\Omega_m$ degeneracy 
(dotted line, $95\%$ confidence).  Also shown are the $95\%$ confidence
regions obtained using all but the lowest six bins (dashed) and all bins
(solid).   This illustrates the fact that probing low 
halo masses breaks the $\sigma_8-\Omega_m$ degeneracy by a considerable
amount.  Therfore, we do not expect to recover the usual cluster normalization
condition since we are probing rather low halo masses.}
\label{low_hmf}
\end{figure}

We begin our analysis by determining the $95\%$ confidence regions in the
$\sigma_8-\Omega_m$ plane obtained using only our 10 richest bins 
(clusters having 35 galaxies or $M\gtrsim 2\cdot 10^{14} M_\odot$)
while holding all parameters except $\sigma_8$ and $\Omega_m$ constant.
This is shown in figure \ref{low_hmf} with a dotted line. The degeneracy
axis in the figure is $\sigma_8\Omega_m^{0.62} = constant$, so
we see then that we do indeed
recover the cluster normalization condition when, a- we hold all other
parameters fixed, and b- restrict ourselves to the most massive clusters. 
This degeneracy simply reflects the fact that the halo mass function at
$m\sim 10^{14} M_\odot$ scales is degenerate according to the above expression
(see e.g. Zheng et al. \cite{Zhengetal}).  On this basis, one would expect
that probing low mass scales would allow us to break this degeneracy, which
is indeed the case.  We show this in figure \ref{low_hmf} where we plot
the confidence regions obtained when including lower mass clusters.
In particular, the confidence regions shown with the dashed and solid
lines in figure \ref{low_hmf} are obtained using
all but the lowest six bins ($N_{gal}>21$ or $M\gtrsim 10^{14} M_\odot$)
and all bins respectively.
We see that with the inclusion of lower mass bins the degeneracy region 
is greatly reduced.  Further, as we probe lower and lower masses, the
axis which is least strongly constrained becomes steeper and steeper,
going from $0.62$ when only the 10 richest bins are used
to $0.73$ when all bins are used.

We can gain further insight on the $\sigma_8-\Omega_m$ degeneracy  
by considering what happens when we allow $M_1$ and $\alpha$ to vary,
but keeping $h$ fixed.
The two most constrained directions are obtained from the estimated
parameter correlation matrix are\footnote{We 
are again neglecting a small dependence on $\sigma_8$ of the form
$\sigma_8^{.03}$ in equation \ref{no2}.}

\begin{eqnarray}
\alpha \Bigl( \sigma_8 \bigl(\Omega_m/M_1\bigr)^{0.67} \Bigr)^{0.50}
       = 0.503\cdot 10^{-12} M_\odot^{-1} \pm 2\%\label{no1} \\
\alpha^{0.32} \bigr(\Omega_m/M_1\bigl)^{-1} = 20.0\cdot 10^{12} M_\odot 
\pm 13\%. \label{no2}
\end{eqnarray}

Note that these two eigenvectors are \it not \rm simply expressions
\ref{egv1} and \ref{egv2} with $h$ held constant.  To see how they
are related, it is best to pretend we know nothing about equations
\ref{egv1} and \ref{egv2} while we analyze the above eigenvectors.
We can then go back and see how the pair of eigenvectors \ref{egv1}
and \ref{egv2} are related to \ref{no1} and \ref{no2}.

Let us then analyze the above eigenvectors, whose 
structure allows for a very simple interpretation. 
Take $\alpha$ and $M_1$ to be fixed.  
Then, the first eigenvector becomes the cluster normalization condition
$\sigma_8\Omega_m^{0.67}$, while the second eigenvector is almost 
entirely $\Omega_m$.  Since the low mass end of the halo mass function
is essentially independent of $\Omega_m$, this suggests the first
eigenvector is driven by the high mass end of the halo mass function only, 
while the second eigenvector is driven by the low mass end alone.  
Indeed, the eigenvectors obtained when restricting ourselves 
to high mass bins (clusters with $35$ galaxies or more) are close
to those above, except that we do not lose any constraining
power for the first eigenvector, while the constraint on the
second eigenvector is weakened by a factor of four.  We conclude then
that equation \ref{no1} may be thought of as the 
constraint arising from matching the high mass end of the halo mass function,
while equation \ref{no2} arises from matching the low mass end.

Are the vectors \ref{no1} and \ref{no2} related to the vectors we found
in section \ref{what_abundances_tell_us}? To approach this question,
we can consider what happens to the vectors \ref{no1} and \ref{no2}
when we let $h$ vary.

Consider first what happens at a qualitative level: $h$ affects both
the high mass end and the low mass end of the halo mass function. As
such, we do not expect our eigenvectors to cleanly separate into 
matching one or the other end of the halo mass functions- they will 
mix.  Indeed, when we let $h$ vary, the first eigenvector becomes

\begin{equation}
\alpha\Bigl(\sigma_8 \bigl(\Omega_m/M_1\bigr)^\gamma\Bigr)^{0.50} 
	h^{\tilde\gamma} = constant
\label{gencon}
\end{equation}

where $\gamma$ and $\tilde \gamma$ depend on the lowest mass scale
probed. e.g. $\gamma=0.73$ when all bins are used (as we
found in section \ref{what_abundances_tell_us}) while
$\gamma=0.62$ when we use only rich clusters.

We note several things: first, the vector \ref{gencon} is essentially
identical to $\ref{egv1}$ when $n$ is held fixed.  Interestingly, though,
the $\sigma_8\Omega_m^\gamma$ degeneracy no longer has the constant
exponent $\gamma\approx 0.62$.  The variation of the exponent comes about 
because of the mixing of the constraints from the low and high mass ends
of the halo mass function, and hence our result above compromises by giving
us the most strongly constrained direction in the $\sigma_8-\Omega_m$
plane when all other parameters are held fixed.

What about the eigenvector from expression \ref{no2}?  
Since $h$ mixes the high and low mass end constrains, one may expect
the eigenvector \ref{no2} not only to be drastically altered 
(since it no longer represent the low mass end constraint), but also
to be greatly weakened since the constraining power of the 
high and low mass ends of the halo mass function now goes into the first 
eigenvector.  This is indeed the case.   
Despite this result, however, there is still a second highly constrained
eigenvector, which is essentially that of expression \ref{egv2} with 
$n$ fixed.  This eigenvector is a new constrain that arises from 
allowing $h$ to vary, and is thus not associated with the eigenvector
from expression \ref{no2}.

\subsection{Hiding the $\sigma_8-\Omega_m$ Degeneracy}

\begin{figure}[t]
\sfig{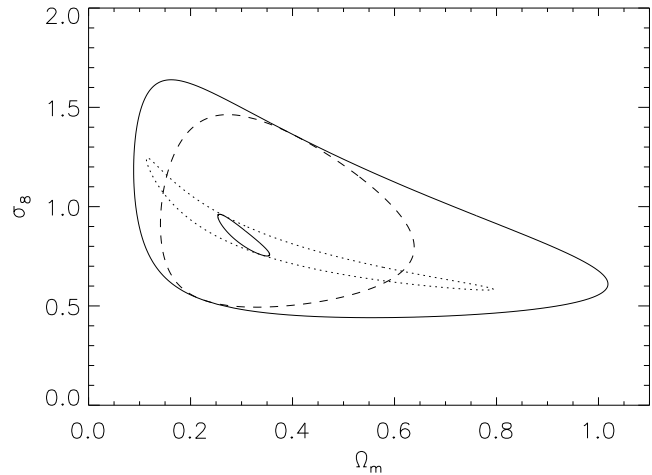}{0.55\columnwidth}
\caption{\footnotesize The effect of allowing various parameters to vary
in determining the $\sigma_8-\Omega_m$ confidence regions is shown above.
All curves above are $95\%$ confidence and hold $n$ and $\Omega_bh^2$
fixed.  The inner solid curve also holds $\alpha,M_1,$ and $h$ fixed.
The dashed curve allows $\alpha$ and $M_1$ to vary with priors
$\Delta\alpha/\alpha =10\%$ and $\Delta M_1/M_1=30\%$.  The dotted curve
holds $\alpha$ and $M_1$ fixed, but varies $h$ with the prior
$\Delta h=0.1$.  Finally, the outer solid curve allows $\alpha,M_1,$
and $h$ to vary with the aforementioned gaussian priors.}
\label{s8_om2}
\end{figure}

We argued throughout the text that the appearance of 
the $\alpha-\sigma_8$ and $M_1-\Omega_m$ degeneracies makes marginalization
over the mass tracer scaling relation necessary if one wishes to
avoid placing overly optimistic constraints on the cosmological parameters.
A by product of this marginalization is that the characteristic shape
of the $\sigma_8-\Omega_m$ degeneracy is then completely washed out.
We illustrate this effect below, not only to observe the degradation of
the confidence regions, but also because  we can get 
a better feel as to how important the various effects are.

We begin with the constraints on $\sigma_8$ and $\Omega_m$ when all other
parameters are fixed.  This is shown in figure \ref{s8_om2} as the 
inner solid ellipse ($95\%$ confidence region), 
which matches that of figure \ref{low_hmf}. 
We now let $\alpha$ and $M_1$ to vary by assuming gaussian priors 
$\Delta\alpha/\alpha=10\%$ and $\Delta M_1/M_1=30\%$. 
 This is shown as a dashed
curve in figure \ref{s8_om2}.  The effect of allowing $\alpha$ and 
$M_1$ to vary is staggering- the confidence regions are enormously expanded.

We can likewise observe the effect of allowing $h$ to vary while keeping 
the other parameters fixed.  Figure \ref{s8_om2} shows the $95\%$ confidence
regions (dotted line) 
marginalized over $h$ with a gaussian prior $\Delta h=0.1$.  As 
we expect, one direction remains tightly constrained, corresponding to
the eigenvector \ref{egv2}, the generalized cluster abundance normalization
condition.  On the other hand, the perpendicular direction is weakened to
a large degree, again reflecting that a floating $h$ mixes the
high and low mass ends constraints from the halo mass function into 
a single constraint.

Finally, shown with the outer solid curve 
in figure \ref{s8_om2} is the $95\%$ confidence
region when $\alpha,M_1,$ and $h$ are allowed to vary using the above
priors.  This last curve is the true constraint one may expect from
cluster abundances alone.  Also shown as reference with the thicker
solid curve is the $95\%$ confidence region obtained including bias 
information.  All priors for this last curve are the same, 
except for $\alpha$, for which no prior was assumed.

\end{document}